\documentclass{amsart}

\usepackage{amsmath}
\usepackage{amssymb}
\usepackage[mathscr]{eucal}
\usepackage{graphicx}
\usepackage{enumerate}
\usepackage{hyperref}

\usepackage[english]{babel}

\theoremstyle{plain} 
\newtheorem{theorem}{Theorem}[section]
\newtheorem{proposition}[theorem]{Proposition}
\newtheorem{lemma}[theorem]{Lemma}
\newtheorem{corollary}[theorem]{Corollary}

\newtheorem{remark}[theorem]{Remark}

\newcommand{\Lemref}[1]{Lemma~\ref{#1}}
\newcommand{\Corref}[1]{Corollary~\ref{#1}}
\newcommand{\Propref}[1]{Proposition~\ref{#1}}
\newcommand{\Thmref}[1]{Theorem~\ref{#1}}

\newcommand{\Sectionref}[1]{Section~\ref{#1}}

\newcommand{\reals}{{\mathbb R}}
\newcommand{\from}{\colon}
\DeclareMathOperator{\with}{:}
\DeclareMathOperator{\Order}{\mathcal{O}}
\newcommand{\Lp}[1]{L^{#1}}
\newcommand{\LpMu}[2]{L^{#1}_{#2}}
\newcommand{\bydef}{{:=}}
\newcommand{\scalar}[2]{{\left\langle\, #1, #2\,\right\rangle}}
\newcommand{\norm}[1]{\left\Arrowvert \, #1 \, \right\Arrowvert}
\DeclareMathOperator{\Span}{span}
\newcommand{\diam}{{\mathrm{diam}\,}}
\newcommand{\const}{{\mathrm{const}}}

\DeclareMathOperator{\expect}{{\mathbb E}}
\newcommand{\TVnorm}[1]{|#1|_{\mathrm{TV}}}
\DeclareMathOperator{\given}{|}
\DeclareMathOperator{\var}{Var}
\DeclareMathOperator{\Min}{\wedge}
\DeclareMathOperator{\Max}{\vee}

\newcommand{\unitVectorI}[1]{\mathbf{e}_{#1}}
\newcommand{\vecOne}{\mathbf{1}}
\newcommand{\contractionMatrixN}[1]{\mathcal{C}^{(#1)}}

\newcommand{\spectrum}[1]{\sigma(#1)}
\newcommand{\spectrumE}[1]{E(#1)}

\newcommand{\EnergyDist}{\mu}

\newcommand{\MarkovChainEnergySiteRatiodist}{\nu_{r}}
\newcommand{\MarkovChainEnergySiteSumdist}{\nu_s}
\newcommand{\MarkovChainEnergySitedist}{\nu}

\newcommand{\transitionP}{P}
\newcommand{\rateFunction}{\Lambda}
\newcommand{\rateSum}{\Lambda_s}
\newcommand{\rateRatio}{\Lambda_r}

\newcommand{\varPstar}{\sigma_{P^\star}^2}
\newcommand{\varP}{\sigma_P^2}
\newcommand{\MarkovChainEnergydist}{\mu}
\newcommand{\MarkovChainUdist}{\mu}
\newcommand{\MarkovChainUdistT}[1]{\mu_{#1}}
\newcommand{\MarkovChainUudist}{\mu'}
\newcommand{\MarkovChainUudistT}[1]{\mu'_{#1}}

\newcommand{\MarkovChainEnergyI}[1]{\mathsf{X}_{#1}}
\newcommand{\MarkovChainEnergyT}[1]{\mathsf{X}(#1)}
\newcommand{\MarkovChainEnergyIT}[2]{\mathsf{X}_{#1}(#2)}
\newcommand{\MarkovChainEEnergyT}[1]{\mathsf{X}'(#1)}
\newcommand{\EnergyState}{x}
\newcommand{\EnergyStateI}[1]{x_{#1}}
\newcommand{\EEnergyState}{x'}
\newcommand{\EEnergyStateI}[1]{x'_{#1}}
\newcommand{\MarkovChainU}{\mathsf{U}}
\newcommand{\MarkovChainUT}[1]{\mathsf{U}(#1)}
\newcommand{\MarkovChainUTI}[2]{\mathsf{U}_{#2}(#1)}
\newcommand{\MarkovChainUState}{u}
\newcommand{\MarkovChainUStateI}[1]{u_{#1}}
\newcommand{\MarkovGeneratorU}{\hat{\mathcal{L}}}
\newcommand{\MarkovChainUu}{\mathsf{U}'}
\newcommand{\MarkovChainUuT}[1]{\mathsf{U}'(#1)}
\newcommand{\MarkovChainUuState}{u'}
\newcommand{\MarkovChainUuStateI}[1]{u'_{#1}}
\newcommand{\MarkovGeneratorUU}{\bar{\mathcal{L}}}

\newcommand{\mapTUIA}[2]{\hat{T}^{\epsilon}_{#1,#2}}
\newcommand{\mapTIA}[2]{T_{#1,#2}}

\newcommand{\VasersteinDistSymbol}{\rho}
\newcommand{\VasersteinDistp}[3]{\VasersteinDistSymbol_{#1}(#2, #3)}
\newcommand{\VasersteinDist}[2]{\VasersteinDistSymbol(#1, #2)}
\newcommand{\distUSymbol}{\hat{\mathrm{d}}}
\newcommand{\distU}[2]{\distUSymbol(#1, #2)}
\newcommand{\distSymbol}{\mathrm{d}}
\newcommand{\dist}[2]{\distSymbol(#1, #2)}
\newcommand{\Lip}[1]{\mathrm{Lip}(#1)}

\newcommand{\MarkovDirichletForm}{\mathcal{D}}
\newcommand{\MarkovGenerator}{\mathcal{L}}

\newcommand{\stateSpaceU}{\hat{{\mathcal S}}}
\newcommand{\stateSpace}{{\mathcal S}}
\newcommand{\stationaryDist}{\pi}
\newcommand{\stationaryDistEN}[2]{\pi_{#1,#2}}
\newcommand{\transitionOperator}{{\mathscr P}}

\begin{document}

\title{Mixing rates of particle systems with energy exchange}

\author{Alexander Grigo}
\author{Konstantin Khanin}
\author{Domokos Szasz}
\address{
Department of Mathematics, University of Toronto, Toronto, ON, Canada.
}
\email{alexander.grigo@utoronto.ca}
\address{
Department of Mathematics, University of Toronto, Toronto, ON, Canada.
}
\email{khanin@math.utoronto.ca}
\thanks{
  A.G. gratefully acknowledges the generous support and warm hospitality
  of the Fields Institute and the
  Mathematical Institute of the
  Budapest University of Technology and Economics.
}
\address{
Budapest University of Technology and Economics,
Mathematical Institute, Egry J. u. 1, 1111 Budapest, Hungary.
}
\email{szasz@math.bme.hu}
\thanks{
  D. Sz. expresses his sincere thanks to
  Department of Mathematics of University of Toronto and to the Fields
  Institute, Toronto where part of this work has been done (Fall term 2009 and
  June 2011, resp.),
and is grateful to Hungarian National Foundation for Scientific
Research grants No. T 046187, K 71693,
NK 63066 and TS 049835.
}


\date\today

\maketitle

\begin{abstract}
  A fundamental problem of non-equilibrium statistical mechanics
  is the derivation of macroscopic transport equations in the
  hydrodynamic limit. The rigorous study of such limits
  requires detailed information
  about rates of convergence to equilibrium for finite sized systems.
  In this paper we consider the finite
  lattice $\{1, 2, \ldots, N\}$, with an energy
  $\EnergyStateI{i} \in (0,\infty)$ associated to each site.
  The energies evolve according to a Markov jump process
  with nearest neighbor interaction
  such that the total energy is preserved.

  We prove that for an entire class of such models
  the spectral gap of the generator of the Markov process
  scales as $\Order(N^{-2})$.
  Furthermore, we provide a complete classification of reversible
  stationary distributions of product type.
  We demonstrate that our results apply to models similar
  to the billiard lattice model
  considered in \cite{10863485},
  and hence provide a first step in
  the derivation of a macroscopic heat equation for a microscopic
  stochastic evolution of mechanical origin.
\end{abstract}


\section{Introduction}

\subsection{Motivation and related works}

A fundamental problem of non-equilibrium statistical mechanics
is the derivation of effective equations in the
hydrodynamic limit.
Often these are hydrodynamic equations (Euler, Navier-Stokes),
or related transport equations (Burgers equation, heat equation).
There are very few models for which rigorous results exist. They include
particle models like simple exclusion, zero range processes,
see \cite{MR1707314} and references therein, and
continuous systems like the Ginzburg-Landau equation
\cite{MR914986,MR894407,MR954674} and the model of \cite{MR656869}.

The rigorous study of hydrodynamic limits
requires detailed information
about rates of convergence to equilibrium for finite sized systems,
especially if the system is of non-gradient type.
In particular, the scaling of the spectral gap of the generator
with the system size $N$ is of crucial importance.
Such information is typically obtained
by analyzing the Dirichlet form, corresponding to the explicitly
known stationary distributions.

Obtaining good estimates (in terms of the system size)
on the spectral gap of the generator
is highly non-trivial. For example, to obtain the corresponding results for the
Kac model \cite{MR0084985} it took almost half a century
\cite{MR1825150} (using Yau's martingale method \cite{MR1414837,MR1483598})
and
\cite{MR2020418,MR2403324}.

Recently there has been a growing interest in and hope for
establishing hydrodynamic
limits for systems that are either purely deterministic
or originate (somehow) from deterministic, in particular mechanical, models.
A program to obtain information about the stationary distributions under
the influence of stochastic boundary conditions was proposed in
\cite{MR2200889}. Another approach was suggested in the recent series
of papers \cite{10297039,10863485}, where the analysis of
the hydrodynamic limit of a billiards lattice model was
outlined by following a two-step procedure. In the first step
the deterministic dynamics is rescaled in order to obtain a
mesoscopic stochastic model (also referred to as master equation).
In a second step the hydrodynamic behavior of the
mesoscopic stochastic model should be derived.

For neither of the two steps proposed in \cite{10297039,10863485}
rigorous results are available.
Deriving master equations from interacting
mechanical models is a very difficult problem. Only recently
some rigorous results in this direction were obtained in
\cite{2010arXiv1010.3972D},
where the weak interaction limit is considered opposed
to the rare interaction limit of \cite{10297039,10863485}.
As a matter of fact, the second step, i.e.
deriving the hydrodynamic limit from the master equation, seems
a much more tractable mathematical problem.
The present paper is an attempt to make a first step in this direction
by providing information about the spectral gap of the generator
of an entire class of models, which are of similar type as
the master equation of the billiard lattice model considered
in \cite{10297039,10863485}. In particular,
the model \cite{10863485} belongs to the class of models we are
considering, the obtained spectral bound is exactly the necessary one
for which the derivation of the hydrodynamics limit is feasible.

\subsection{Description of the model}

The model we consider in this paper is as follows.
Let $N\geq 2$ be an integer, and consider the
lattice $\{1, 2, \ldots, N\}$. To every site $i$
of this lattice we associate an energy $\EnergyStateI{i}$,
which is a positive real number. The collection of all the
energies will be denoted by
$
\EnergyState
=
(\EnergyStateI{1}, \ldots, \EnergyStateI{N})
\in \reals_+^N
$.
To each nearest neighbor pair of the lattice we
associate an independent exponential clock with
a rate $\rateFunction$ that depends on the total energy of this pair.
As soon as one of the $N-1$ clocks rings, say for the
pair $(i, i+1)$, then a number $0\leq \alpha \leq 1$ is
drawn, independently of everything else, according to a distribution
$\transitionP$, that only depends on the two energies $\EnergyStateI{i}$,
$\EnergyStateI{i+1}$. The update of the energies is then such that
the new energy at site $i$
is $\alpha\,(\EnergyStateI{i} + \EnergyStateI{i+1})$,
the new energy at site $i+1$
is $(1-\alpha)\,(\EnergyStateI{i} + \EnergyStateI{i+1})$,
and all other energies remain unchanged.

This procedure defines a continuous time Markov jump process
$\MarkovChainEnergyT{t}$ on $\reals_+^N$. More formally, we define
the process $\MarkovChainEnergyT{t}$ by its
infinitesimal generator $\MarkovGenerator$, acting on
bounded
\footnote{
Throughout this paper we will always assume that the various functions
are Borel measurable without stating this assumption explicitly. This
will not lead to confusion, since higher regularity assumptions (like
continuity or integrability) are stated explicitly.
}
functions $A \from \reals_+^N \to \reals$ as
\begin{equation}
  \label{eqn_definitionGenerator}
  \begin{split}
    \MarkovGenerator A(\EnergyState)
    &=
    \sum_{i=1}^{N-1}
    \rateFunction( \EnergyStateI{i}, \EnergyStateI{i+1} )
    \int \transitionP(\EnergyStateI{i}, \EnergyStateI{i+1} ,d\alpha )
    \,[A( \mapTIA{i}{\alpha} \EnergyState)
    -
    A(\EnergyState) ]
  \end{split}
\end{equation}
where $\rateFunction \from \reals_+^2 \to \reals_+$
is continuous, and
$\transitionP(\EnergyStateI{i}, \EnergyStateI{i+1} ,d\alpha )$
is a probability measure on $[0,1]$, which depends continuously on
$(\EnergyStateI{i}, \EnergyStateI{i+1}) \in \reals_+^2$.
The maps $\mapTIA{i}{\alpha}$
model the energy exchange between the neighboring sites $i$ and $i+1$,
and are defined by
\begin{equation}
  \label{eqn_definition_mapT}
  \mapTIA{i}{\alpha}(\EnergyState)
  =
  \EnergyState
  +
  [
  \alpha\,\EnergyStateI{i+1}
  -
  (1-\alpha)\,\EnergyStateI{i}
  ]
  \,[\unitVectorI{i} - \unitVectorI{i+1}]
\end{equation}
where $\unitVectorI{i}$ denotes the $i$-th unit vector of $\reals^N$.

In particular,
the process $\MarkovChainEnergyT{t}$ preserves the total
energy, i.e. for any two times $t_1$ and $t_2$ the identity
$
\sum_{i=1}^N \MarkovChainEnergyIT{i}{t_1}
=
\sum_{i=1}^N \MarkovChainEnergyIT{i}{t_2}
$
holds.
Therefore, we introduce for any $\epsilon>0$
\footnote{
The parameter $\epsilon$ denotes the average energy per site
and should not be thought of as a necessarily small number.
We hope that this does not cause any confusion, even though
it is a common practice to reserve the use of the symbol $\epsilon$
to denote a small number.
}
the sets
\begin{equation*}
  \stateSpace_{\epsilon,N}
  =
  \Big\{
  \EnergyState \in \reals_+^N
  \with
  \sum_{i=1}^N \frac{1}{N}\,\EnergyStateI{i}
  =
  \epsilon
  \Big\}
\end{equation*}
which are invariant for the process $\MarkovChainEnergyT{t}$.
The value of $\epsilon$ represents the mean energy per site.

Since $\stateSpace_{\epsilon,N}$ is compact and invariant the assumed
continuity of $\rateFunction$ and $\transitionP$ guarantees
the existence of at least one stationary distribution
$\stationaryDistEN{\epsilon}{N}$
for $\MarkovChainEnergyT{t}$
on each $\stateSpace_{\epsilon,N}$.
As we pointed out, the scaling of the rate of convergence
towards the stationary distribution in terms of the lattice size $N$
is of crucial importance in studying
the hydrodynamic limit of this model rigorously.

\subsection{Outline of the paper}

The purpose of this paper is to present a dynamical and geometric
approach to establish the scaling of the spectral
gap of the generator
\eqref{eqn_definitionGenerator}
under rather general assumptions on the rate function $\rateFunction$
and transition kernel $\transitionP$.
The strategy we adopt is as follows.
In \Sectionref{sect_gap_generalCase}
we show that for a large class of
rates $\rateFunction$ and transition operators $\transitionP$ the
scaling of the spectral gap of the corresponding generator
\eqref{eqn_definitionGenerator}
can be obtained by considering only the special case
of a constant rate $\rateFunction$
and a state independent transition kernel $\transitionP$.
The precise statement is formulated in \Thmref{thm_comparisionTheorem},
which we prove under the
two key assumptions: the reversibility
of the process $\MarkovChainEnergyT{t}$, and the existence of
a lower bound on the rate function $\rateFunction$.
The requirement of a lower bound on the rate function
seems to be a technical condition, but it cannot be
removed at present.

In \Sectionref{section_examples} we show that
(a slight modification of) the three-dimensional
stochastic billiard lattice model of \cite{10863485}
is a special case of the general model considered in the
present paper,
provided that one introduces a lower cut-off for the rate function
originally considered in \cite{10863485}.
In particular, we show that it then follows that
the spectral gap scales as $\Order(N^{-2})$.

Since we assume reversibility of the
stationary distribution to derive the spectral properties, we provide in
\Sectionref{sect_classification_products}
a classification
of reversible stationary distributions of product type. Such
measures are of particular interest in the hydrodynamic limit, and
appear naturally in mechanical models and statistical mechanics in form
of Gibbs measures.
We show in \Thmref{thm_reversibleProductMeasures}
that if a model of the class
\eqref{eqn_definitionGenerator}
considered in this paper
admits a reversible product distributions, then
this measure must necessarily be a product Gamma-distributions
(or a single atom). This is precisely the type of product measures
considered in statistical mechanics for mechanical models.

The main part of the paper deals with establishing the scaling of the
spectral gap of the generator for the process with constant rates
$\rateFunction$ and state independent transition kernel $\transitionP$.
This case is studied in \Sectionref{section_reference}.
The key difference of our analysis when
compared to the above mentioned related works is that
instead of focusing directly on $\Lp{2}$ convergence, for example
by analyzing the associated Dirichlet form,
we first establish weak convergence towards
a stationary distribution. For the later part it is crucial that
this weak convergence is made quantitative in a sufficiently strong
metric for the weak topology. For this purpose we use the Vaserstein
distance and prove in \Thmref{thm_ergodicity} that there is an
exponential rate of convergence of $\MarkovChainEnergyT{t}$
to equilibrium, which scales as
$\Order(N^{-2})$ in the system size $N$.
The key step in the proof is to construct an adapted metric on the
state space of $\MarkovChainEnergyT{t}$, for which the contraction
property can be established. This requires special coordinates and
a coupling argument, which is presented in
\Propref{prop_average_contraction} and
\Propref{prop_convergence_vaserstein}.

The advantage of first establishing exponential convergence in the
weak sense is that it allows to include very general transition
kernels $\transitionP$ (for example, non-absolutely continuous kernels),
and does make reference to the invariant measure. Instead
it relies on a very natural geometric property of the interaction
mechanism of $\MarkovChainEnergyT{t}$.

In a second step we assume reversibility of the constructed unique invariant
measure, and show that the $\Lp{2}$ convergence occurs at an exponential
rate, which is explicitly related to the rate of convergence in
Vaserstein metric. In particular, this shows that the spectral gap
scales as $\Order(N^{-2})$ in the lattice size $N$.
The precise statement is given in
\Thmref{thm_spectrumL2_reversible_constantRates}, whose prove
relies on the Kantorovich-Rubinstein duality property of the Vaserstein
metric, see \Lemref{lem_LipschitzContraction}. This is another
manifestation of the
usefulness of the weak convergence in Vaserstein distance in
the study of the spectral gap for interacting particle systems.

\Sectionref{sect_conclusion} contains final comments and conclusions.

\section{Analysis of a special case}
\label{section_reference}

In this section we consider a special case of
the class of processes defined by generators of the
form \eqref{eqn_definitionGenerator}. Namely we consider
the case where
the rate function $\Lambda$ is constant, and
the transition kernel $P$ is state independent.
In other words we consider a process $\MarkovChainEnergyT{t}$
with infinitesimal generator
\begin{equation}
  \label{eqn_definitionGenerator_reference}
  \MarkovGenerator A(\EnergyState)
  =
  \Lambda
  \sum_{i=1}^{N-1}
  \int P(d\alpha)
  \,[A( \mapTIA{i}{\alpha} \EnergyState)
  -
  A(\EnergyState) ]
\end{equation}
acting on the space of bounded observables $A\from \reals_+^N \to \reals$.

As was already mentioned the process $\MarkovChainEnergyT{t}$
preserves the total energy. This implies that
the process cannot have a unique stationary state on all
of $\reals_+^N$. However, we will show below that the restriction
of the process to any of the invariant sets $\stateSpace_{\epsilon,N}$
has a unique stationary distribution.

The first step in this direction is to introduce more convenient
coordinates on $\stateSpace_{\epsilon,N}$, which is the purpose
of the next result.
\begin{lemma}[$\EnergyState$ in terms of $\MarkovChainUState$]
  \label{lem_localCoordinates_x_u}
  Let $N$ and $\epsilon$ be fixed. Then any
  $\EnergyState \in \stateSpace_{\epsilon,N}$
  can be uniquely written as
  \begin{align*}
    \EnergyState
    &=
    \epsilon\,\vecOne
    +
    \sum_{i=1}^{N-1}
    \MarkovChainUStateI{i}\,[\unitVectorI{i} - \unitVectorI{i+1}]
  \end{align*}
  for some $\MarkovChainUState \in \reals^{N-1}$, where
  $\vecOne$ denote the vector $(1, \ldots, 1)$.
  Furthermore, via this change
  of coordinates the set
  $\stateSpace_{\epsilon,N} \subset \reals_+^N$ is in one-to-one
  correspondence with the set
  \begin{equation*}
    \stateSpaceU_{\epsilon,N}
    =
    \{
    \MarkovChainUState \in \reals^{N-1}
    \with
    -\epsilon \leq \MarkovChainUStateI{1}
    ,\;
    \MarkovChainUStateI{i-1} \leq \epsilon + \MarkovChainUStateI{i}
    ,\;
    \MarkovChainUStateI{N-1} \leq \epsilon
    \}
    \;.
  \end{equation*}
\end{lemma}

Note that the vectors $\unitVectorI{i} - \unitVectorI{i+1}$
for $i=1,\ldots,N-1$ span the simplex $\stateSpace_{\epsilon,N}$,
but they are not mutually orthogonal. However, they almost are
in the sense that any two of them are perpendicular as soon as
they correspond to two values of $i$, which differ by at least $2$.

In the following we will also need the inverse coordinate transformation,
which expresses $\MarkovChainUState$ in terms of $\EnergyState$.
\begin{lemma}[$\MarkovChainUState$ in terms of $\EnergyState$]
  \label{lem_T_localCoordinates_static}
  Let $\EnergyState \in \reals_+^N$ be given.
  Then the corresponding $\epsilon$ is given by
  $
  \epsilon = \sum_{i=1}^N \frac{1}{N}\,\EnergyStateI{i}
  $,
  and the corresponding $\MarkovChainUState$ is the solution to the discrete
  Poisson equation with Dirichlet boundary conditions
  \begin{align*}
    \MarkovChainUStateI{i-1}
    - 2\,\MarkovChainUStateI{i}
    + \MarkovChainUStateI{i+1}
    =
    \EnergyStateI{i+1}
    -
    \EnergyStateI{i}
    \qquad\text{for}\qquad
    i=1,\ldots,N-1
  \end{align*}
  where we formally set
  $\MarkovChainUStateI{0} \equiv \MarkovChainUStateI{N} \equiv 0$.
  More explicitly
  \begin{equation*}
    \MarkovChainUStateI{i}
    =
    \sum_{k=1}^i (\EnergyStateI{k} - \epsilon)
    =
    \Big[ 1- \frac{i}{N}\Big] \sum_{k=1}^i \EnergyStateI{k}
    -
    \frac{i}{N}\sum_{k=i+1}^N \EnergyStateI{k}
    \qquad\text{for all}\qquad 1 \leq i \leq N-1
  \end{equation*}
  is the expression for the corresponding $\MarkovChainUState\in\reals^{N-1}$.
\end{lemma}
\begin{proof}
  Clearly, $\EnergyState \in \stateSpace_{\epsilon,N}$
  if and only if $\epsilon$ is given by the claimed formula.
  Furthermore, it follows immediately from the definition of
  the coordinates $\MarkovChainUState$, that
  $
  \EnergyStateI{i}
  =
  \epsilon
  + \MarkovChainUStateI{i}
  - \MarkovChainUStateI{i-1}
  $
  for all $i$, where we use the convention
  $\MarkovChainUStateI{0} \equiv \MarkovChainUStateI{N} \equiv 0$.
  This implies that $\MarkovChainUState$ must solve the discrete Poisson
  equation with zero Dirichlet boundary conditions.

  On the other hand we can sum up the expression for
  $\EnergyStateI{i}$ in terms of $\MarkovChainUState$
  and obtain a telescoping sum, which yields
  \begin{align*}
    \MarkovChainUStateI{i}
    =
    \sum_{k=1}^i (\MarkovChainUStateI{k} - \MarkovChainUStateI{k-1})
    =
    \sum_{k=1}^i (\EnergyStateI{k} - \epsilon)
  \end{align*}
  for all $i = 1, \ldots, N-1$.

  And since $\epsilon\,N = \sum_{i=1}^N \EnergyStateI{i}$
  we can replace
  $\epsilon$ in terms of this sum, and thus obtain the
  second expression for $\MarkovChainUStateI{i}$.
\end{proof}

The point of the change of coordinates from $\EnergyState$
to $\epsilon$ and $\MarkovChainUState$ is to separate out
the conserved quantity $\epsilon$, and consider only the evolution
of the nontrivial part $\MarkovChainUT{t}$ of
the process $\MarkovChainEnergyT{t}$
\begin{equation}
  \label{eqn_definition_U}
  \MarkovChainEnergyT{t}
  =
  \epsilon\,\vecOne
  +
  \sum_{i=1}^{N-1}
  \MarkovChainUTI{t}{i}\,[\unitVectorI{i} - \unitVectorI{i+1}]
  \;,
\end{equation}
namely the $\MarkovChainUState$-coordinate vector corresponding to
$\MarkovChainEnergyT{t}$.
Since $\epsilon$ is conserved it follows that $\MarkovChainUT{t}$
is also a homogeneous Markov process (for each $\epsilon$ separately).
Using the results of
\Lemref{lem_localCoordinates_x_u}
and
\Lemref{lem_T_localCoordinates_static}
we can now derive the infinitesimal generator of $\MarkovChainUT{t}$.

\begin{lemma}[The generator of $\MarkovChainUT{t}$]
  \label{lem_T_localCoordinates}
  Let $N$ and $\epsilon$ be fixed. Then the process
  $\MarkovChainUT{t}$ is a homogeneous Markov process
  on $\stateSpaceU_{\epsilon,N}$,
  whose infinitesimal generator $\MarkovGeneratorU_{\epsilon,N}$
  is given by
  \begin{align*}
    \MarkovGeneratorU_{\epsilon,N} A(\MarkovChainUState)
    &=
    \Lambda
    \sum_{i=1}^{N-1}
    \int P(d\alpha)\,[
    A(\mapTUIA{i}{\alpha} \MarkovChainUState)
    -
    A(\MarkovChainUState)
    ]
    \;,
  \end{align*}
  where
  \begin{align*}
    \mapTUIA{i}{\alpha} \MarkovChainUState - \MarkovChainUState
    &=
    [
    (1-\alpha)\,\MarkovChainUStateI{i-1}
    +
    \alpha\,\MarkovChainUStateI{i+1}
    +
    (2\,\alpha-1)\,\epsilon
    -
    \MarkovChainUStateI{i}
    ]
    \,\unitVectorI{i}
    \in \reals^{N-1}
  \end{align*}
  with the convention
  $\MarkovChainUStateI{0} \equiv \MarkovChainUStateI{N} \equiv 0$.
\end{lemma}
\begin{proof}
  From its definition \eqref{eqn_definition_mapT}
  we have
  $
  \mapTIA{i}{\alpha}(\EnergyState)
  =
  \EnergyState
  +
  [
  \alpha\,\EnergyStateI{i+1}
  -
  (1-\alpha)\,\EnergyStateI{i}
  ]
  \,[\unitVectorI{i} - \unitVectorI{i+1}]
  $.
  Note that $[\mapTIA{i}{\alpha}  \EnergyState]_k$ agrees
  with $\EnergyStateI{k}$ for all $k$ different from $i$ and $i+1$, and
  $
  [\mapTIA{i}{\alpha}  \EnergyState]_i
  +
  [\mapTIA{i}{\alpha}  \EnergyState]_{i+1}
  $
  equals
  $\EnergyStateI{i} + \EnergyStateI{i+1}$
  (local energy conservation).
  Therefore, $[\mapTUIA{i}{\alpha} \MarkovChainUState]_k$
  equals $\MarkovChainUStateI{k}$ for all $k \neq i$, because
  by \Lemref{lem_T_localCoordinates_static}
  we have
  $
  \MarkovChainUStateI{i}
  =
  \sum_{k=1}^i (\EnergyStateI{k} - \epsilon)
  $.

  So it remains to consider
  $[\mapTUIA{i}{\alpha} \MarkovChainUState]_i$.
  Using the above two expressions for
  $ \MarkovChainUState $
  and
  $\mapTIA{i}{\alpha}(\EnergyState)$ we obtain
  \begin{align*}
    [\mapTUIA{i}{\alpha} \MarkovChainUState]_i
    -
    \MarkovChainUStateI{i}
    &=
    \sum_{k=1}^i (
    [\mapTIA{i}{\alpha}(\EnergyState)]_k
    - \epsilon)
    -
    \sum_{k=1}^i ( \EnergyStateI{k} - \epsilon)
    =
    [\mapTIA{i}{\alpha}(\EnergyState)]_i - \EnergyStateI{i}
    \\
    &=
    \alpha\,\EnergyStateI{i+1} - (1-\alpha)\,\EnergyStateI{i}
    \;.
  \end{align*}
  Using \Lemref{lem_localCoordinates_x_u} we can
  express $\EnergyState$ in terms of $\MarkovChainUState$
  as
  $
  \EnergyStateI{i}
  =
  \epsilon + \MarkovChainUStateI{i} - \MarkovChainUStateI{i-1}
  $, where we used the convention
  $\MarkovChainUStateI{0} \equiv \MarkovChainUStateI{N} \equiv 0$.
  Substituting this expression in the previous formula yields
  the claimed expression for
  $
  \mapTUIA{i}{\alpha} \MarkovChainUState
  -
  \MarkovChainUState
  $.
  Furthermore, this (trivially) also shows the claimed expression for the
  infinitesimal generator of $\MarkovChainUT{t}$.
\end{proof}

\subsection{Weak convergence}

Fix again the values of $\epsilon$ and $N$.
To study the existence of and rate of convergence to a stationary
distribution we consider a bivariate Markov process
$(\MarkovChainUT{t}, \MarkovChainUuT{t})$
on
$\stateSpaceU_{\epsilon,N} \times \stateSpaceU_{\epsilon,N}$,
whose infinitesimal
generator
\begin{equation}
  \label{eqn_definitionGenerator_reference_bivariate}
  \MarkovGeneratorUU A(\MarkovChainUState, \MarkovChainUuState)
  =
  \Lambda
  \sum_{i=1}^{N-1}
  \int P(d\alpha)
  \,[A(\mapTUIA{i}{\alpha} \MarkovChainUState,
  \mapTUIA{i}{\alpha} \MarkovChainUuState)
  -
  A(\MarkovChainUState, \MarkovChainUuState) ]
\end{equation}
for any (bounded) observable $A$ on
$\stateSpaceU_{\epsilon,N} \times \stateSpaceU_{\epsilon,N}$.
Note that this is a special Markov coupling
of two copies of the Markov chains
generated by $\MarkovGeneratorU$.

In order to analyze the weak convergence of the process
$\MarkovChainEnergyT{t}$ towards a stationary distribution
we consider the Vaserstein metric on the probability measures
on $\stateSpace_{\epsilon,N}$. This requires, however, a
metric $\dist{.}{.}$ on $\stateSpace_{\epsilon,N}$.
We equip $\stateSpaceU_{\epsilon,N}$ with the Euclidean metric
\begin{subequations}
  \begin{equation}
    \distU{ \MarkovChainUState }{ \MarkovChainUuState }
    \bydef
    \Big[
    \sum_{i=1}^{N-1}
    ( \MarkovChainUStateI{i} - \MarkovChainUuStateI{i})^2
    \Big]^{\frac{1}{2}}
  \end{equation}
  which corresponds to the metric
  \begin{equation}
    \dist{ \EnergyState }{ \EEnergyState }
    =
    \Big[
    \sum_{i=1}^{N-1}
    \Big(
    \sum_{k=1}^i
    [
    \EnergyStateI{k} - \EEnergyStateI{k}
    ]
    \Big)^2
    \Big]^{\frac{1}{2}}
    \equiv
    \distU{ \MarkovChainUState }{ \MarkovChainUuState }
  \end{equation}
\end{subequations}
on $\stateSpace_{\epsilon,N}$.
In particular, we the have following estimate on the diameter of
$\stateSpace_{\epsilon,N}$.
\begin{lemma}[Diameter of $\stateSpace_{\epsilon,N}$]
  \label{lem_diameter_stateSpace}
  Let $\epsilon$ and $N$ be fixed. Then
  \begin{align*}
    \max_{\EnergyState,\EEnergyState \in \stateSpace_{\epsilon,N}}
    \dist{ \EnergyState }{ \EEnergyState }
    =
    \max_{\MarkovChainUState,\MarkovChainUuState \in \stateSpaceU_{\epsilon,N}}
    \distU{ \MarkovChainUState }{ \MarkovChainUuState }
    \leq
    \epsilon\,N\,\sqrt{N-1}
  \end{align*}
  holds.
\end{lemma}
\begin{proof}
  By \Lemref{lem_localCoordinates_x_u} it follows that
  for any $\MarkovChainUState \in \stateSpaceU_{\epsilon,N}$
  the inequality
  $ -i\,\epsilon \leq \MarkovChainUStateI{i} \leq \epsilon\,(N-i)$
  holds for all $i = 1,\ldots,N-1$. Therefore,
  \begin{equation*}
    \distU{ \MarkovChainUState }{ \MarkovChainUuState }^2
    =
    \sum_{i=1}^{N-1}
    ( \MarkovChainUStateI{i} - \MarkovChainUuStateI{i})^2
    \leq
    \sum_{i=1}^{N-1} ( N\,\epsilon )^2
    =
    \epsilon^2\,N^2\,(N-1)
  \end{equation*}
  for any two $\MarkovChainUState$ and $\MarkovChainUuState$,
  which implies the claim.
\end{proof}

The following \Propref{prop_average_contraction}
provides the first step to estimate
$\distU{ \MarkovChainUT{t} }{ \MarkovChainUuT{t} }$.
A particular role will be played by the matrix
\begin{equation}
  \label{eqn_definition_contractionMatrix}
  \contractionMatrixN{N}
  =
  \begin{pmatrix}
    2 &\vline& 0 & -1 & 0 & 0 &\vline&  \\ \hline
    0 &\vline& 2 & 0 & -1 & 0 &\vline&  \\
    -1 &\vline& 0 & 2 & 0 & -1 &\vline&  \\
    &\vline& &\ddots & &   &\vline& \\
    &\vline& -1 & 0 & 2 & 0 &\vline& -1 \\
    &\vline& 0 & -1 & 0 & 2 &\vline& 0 \\ \hline
    &\vline& 0 & 0 & -1 & 0 &\vline& 2
  \end{pmatrix}
  \in
  \reals^{N \times N}
  \;.
\end{equation}
The spectral properties of $\contractionMatrixN{N-1}$
are provided by the following \Lemref{lem_spectrum_contractionMatrix}.

\begin{lemma}[Spectrum of $\contractionMatrixN{N-1}$]
  \label{lem_spectrum_contractionMatrix}
  If $N$ is odd, then the eigenvalues of
  $\contractionMatrixN{N-1}$ are given by
  \begin{align*}
    4\,\sin^2\Big[ \frac{\pi\,k}{N+1} \Big]
    \qquad\text{for}\qquad
    k=1,\ldots,\frac{N-1}{2}
  \end{align*}
  where each has multiplicity two.
  If $N$ is even, then the eigenvalues of
  $\contractionMatrixN{N-1}$ are given by
  \begin{align*}
    4\,\sin^2\Big[ \frac{\pi\,k}{N} \Big]
    &\qquad\text{for}\qquad
    k=1,\ldots,\frac{N}{2}-1
    \\
    4\,\sin^2\Big[ \frac{\pi\,k}{N+2} \Big]
    &\qquad\text{for}\qquad
    k=1,\ldots,\frac{N}{2}
  \end{align*}
  each of multiplicity one.
\end{lemma}
\begin{proof}
  By the definition of
  $\contractionMatrixN{N}$
  we see that the even and odd indices separate. In fact, it
  is readily seen that the action of
  $\contractionMatrixN{N}$ on the odd indexed
  $(\MarkovChainUStateI{1}, \MarkovChainUStateI{3}, \ldots)$
  and the even indexed
  $(\MarkovChainUStateI{2}, \MarkovChainUStateI{4}, \ldots)$
  entries of $\MarkovChainUState$ is given by the action
  of the matrix
  \begin{equation*}
    A
    =
    \begin{pmatrix}
      2 &\vline& -1 & 0 & 0 &\vline&  \\ \hline
      -1 &\vline& 2 & -1 & 0 &\vline&  \\
      &\vline& &\ddots &   &\vline& \\
      &\vline& 0 & -1 & 2 &\vline& -1 \\ \hline
      &\vline& 0 & 0 & -1 &\vline& 2
    \end{pmatrix}
    \;.
  \end{equation*}
  It is readily verified that
  if $A \in \reals^{m \times m}$, then
  for $k=1, \ldots, m$ the vectors
  $
  ( \sin[\pi\,k\,\frac{1}{m+1}] , \ldots, \sin[\pi\,k\,\frac{m}{m+1}])
  $
  are eigenvectors of $A$ corresponding to the eigenvalues
  \begin{align*}
    4\,\sin^2\Big[ \frac{\pi\,k}{2\,(m+1)} \Big]
    \qquad\text{for}\qquad
    k=1,\ldots,m
    \;.
  \end{align*}

  If $N$ is odd, say $N=2\,m+1$ for some $m\geq 1$, then there
  are $m$ odd and $m$ even indexed entries in
  $\MarkovChainUState\in\reals^{N-1}$.
  Therefore, the eigenvalues of
  $\contractionMatrixN{2\,m}$ are given by
  \begin{align*}
    4\,\sin^2\Big[ \frac{\pi\,k}{2\,(m+1)} \Big]
    \qquad\text{for}\qquad
    k=1,\ldots,m
  \end{align*}
  where each has multiplicity two.

  If $N$ is even, say $N=2\,m+2$ for some $m\geq 1$, then there
  are $m+1$ odd and $m$ even indexed entries in
  $\MarkovChainUState\in\reals^{N-1}$.
  Therefore, the eigenvalues of
  $\contractionMatrixN{2\,m+1}$ are given by
  \begin{align*}
    4\,\sin^2\Big[ \frac{\pi\,k}{2\,(m+1)} \Big]
    &\qquad\text{for}\qquad
    k=1,\ldots,m
    \\
    4\,\sin^2\Big[ \frac{\pi\,k}{2\,(m+2)} \Big]
    &\qquad\text{for}\qquad
    k=1,\ldots,m+1
  \end{align*}
  where each has multiplicity one.
\end{proof}

\begin{proposition}[Average contraction rate]
  \label{prop_average_contraction}
  Assume that the transition kernel $P$ satisfies
  $\int P(d\alpha)\,\alpha = \frac{1}{2}$.
  Then
  \begin{align*}
    \MarkovGeneratorUU[ \distU{ \MarkovChainUState }{ \MarkovChainUuState }^2 ]
    &\leq
    -
    \Lambda\,[1 - 4\,\varP ]
    \,\sin^2\Big[ \frac{\pi}{N+2} \Big]
    \,\distU{ \MarkovChainUState }{ \MarkovChainUuState }^2
  \end{align*}
  holds for any two states
  $\MarkovChainUState$ and $\MarkovChainUuState$,
  where $\varP$ denotes the variance of $P$.
\end{proposition}
\begin{remark}
  Since $P$ is supported on $[0,1]$ and is assumed to have mean
  $\int P(d\alpha)\,\alpha = \frac{1}{2}$ it follows that
  the variance of $P$ satisfies
  $ 0 \leq 1 - 4\,\varP \leq 1$.
\end{remark}
\begin{proof}[Proof of \Propref{prop_average_contraction}]
  From the definition of the generator
  $\MarkovGeneratorUU$ and the distance $\distU{.}{.}$
  it follows
  \begin{align*}
    \MarkovGeneratorUU \distU{ \MarkovChainUState }{ \MarkovChainUuState }^2
    &=
    \Lambda
    \sum_{i=1}^{N-1}
    \int P(d\alpha)
    \,[
    \distU{\mapTUIA{i}{\alpha} \MarkovChainUState}{
    \mapTUIA{i}{\alpha} \MarkovChainUuState}^2
    -
    \distU{\MarkovChainUState}{\MarkovChainUuState}^2
    ]
  \end{align*}
  and
  \begin{align*}
    \distU{\mapTUIA{i}{\alpha} \MarkovChainUState}{
    \mapTUIA{i}{\alpha} \MarkovChainUuState}^2
    &-
    \distU{\MarkovChainUState}{\MarkovChainUuState}^2
    =
    \sum_{k=1}^{N-1}
    \Big[
    (
    [ \mapTUIA{i}{\alpha} \MarkovChainUState ]_k
    -
    [ \mapTUIA{i}{\alpha} \MarkovChainUuState ]_k
    )^2
    -
    ( \MarkovChainUStateI{k} - \MarkovChainUuStateI{k})^2
    \Big]
    \\
    &=
    \sum_{k=1}^{N-1}
    \Big[
    [ \mapTUIA{i}{\alpha} \MarkovChainUState - \MarkovChainUState ]_k
    -
    [ \mapTUIA{i}{\alpha} \MarkovChainUuState - \MarkovChainUuState ]_k
    \Big]
    \cdot
    \\
    &\qquad\qquad\qquad
    \cdot
    \Big[
    [ \mapTUIA{i}{\alpha} \MarkovChainUState - \MarkovChainUState ]_k
    -
    [ \mapTUIA{i}{\alpha} \MarkovChainUuState - \MarkovChainUuState ]_k
    +
    2\, ( \MarkovChainUStateI{k} - \MarkovChainUuStateI{k})
    \Big]
    \;.
  \end{align*}
  Making use of the explicit expression for
  $\mapTUIA{i}{\alpha} \MarkovChainUState - \MarkovChainUState$
  provided by \Lemref{lem_T_localCoordinates}
  \begin{align*}
    \mapTUIA{i}{\alpha} \MarkovChainUState - \MarkovChainUState
    &=
    [
    (1-\alpha)\,\MarkovChainUStateI{i-1}
    +
    \alpha\,\MarkovChainUStateI{i+1}
    +
    (2\,\alpha-1)\,\epsilon
    -
    \MarkovChainUStateI{i}
    ]
    \,\unitVectorI{i}
  \end{align*}
  the above sum simplifies to
  \begin{align*}
    \distU{\mapTUIA{i}{\alpha} \MarkovChainUState}{
    \mapTUIA{i}{\alpha} \MarkovChainUuState}^2
    &-
    \distU{\MarkovChainUState}{\MarkovChainUuState}^2
    \\
    &=
    \Big[
    (1-\alpha)\,[\MarkovChainUStateI{i-1} - \MarkovChainUuStateI{i-1}]
    +
    \alpha\,[\MarkovChainUStateI{i+1} - \MarkovChainUuStateI{i+1}]
    -
    [\MarkovChainUStateI{i} - \MarkovChainUuStateI{i}]
    \Big]
    \cdot
    \\
    &\qquad\qquad
    \cdot
    \,\Big[
    [ \mapTUIA{i}{\alpha} \MarkovChainUState - \MarkovChainUState ]_i
    -
    [ \mapTUIA{i}{\alpha} \MarkovChainUuState - \MarkovChainUuState ]_i
    +
    2\, ( \MarkovChainUStateI{i} - \MarkovChainUuStateI{i})
    \Big]
    \\
    &=
    \Big[
    (1-\alpha)\,[\MarkovChainUStateI{i-1} - \MarkovChainUuStateI{i-1}]
    +
    \alpha\,[\MarkovChainUStateI{i+1} - \MarkovChainUuStateI{i+1}]
    -
    [\MarkovChainUStateI{i} - \MarkovChainUuStateI{i}]
    \Big]
    \cdot
    \\
    &\qquad\qquad
    \cdot
    \,\Big[
    (1-\alpha)\,[\MarkovChainUStateI{i-1} - \MarkovChainUuStateI{i-1}]
    +
    \alpha\,[\MarkovChainUStateI{i+1} - \MarkovChainUuStateI{i+1}]
    +
    [\MarkovChainUStateI{i} - \MarkovChainUuStateI{i}]
    \Big]
    \\
    &=
    \Big[
    (1-\alpha)\,[\MarkovChainUStateI{i-1} - \MarkovChainUuStateI{i-1}]
    +
    \alpha\,[\MarkovChainUStateI{i+1} - \MarkovChainUuStateI{i+1}]
    \Big]^2
    -
    [\MarkovChainUStateI{i} - \MarkovChainUuStateI{i}]^2
    \\
    &=
    (1-\alpha)^2\,[\MarkovChainUStateI{i-1} - \MarkovChainUuStateI{i-1}]^2
    +
    \alpha^2\,[\MarkovChainUStateI{i+1} - \MarkovChainUuStateI{i+1}]^2
    \\
    &\qquad
    +
    2\,\alpha\,(1-\alpha)
    \,[\MarkovChainUStateI{i-1} - \MarkovChainUuStateI{i-1}]
    \,[\MarkovChainUStateI{i+1} - \MarkovChainUuStateI{i+1}]
    -
    [\MarkovChainUStateI{i} - \MarkovChainUuStateI{i}]^2
  \end{align*}
  which in particular shows that the above expression
  depends only on the difference vector
  $\MarkovChainUState - \MarkovChainUuState$.

  Performing now the sum over $i$ yields
  \begin{align*}
    \sum_{i=1}^{N-1}
    &[
    \distU{\mapTUIA{i}{\alpha} \MarkovChainUState}{
    \mapTUIA{i}{\alpha} \MarkovChainUuState}^2
    -
    \distU{\MarkovChainUState}{\MarkovChainUuState}^2
    ]
    =
    (1-\alpha)^2
    \sum_{i=1}^{N-2} [\MarkovChainUStateI{i} - \MarkovChainUuStateI{i}]^2
    +
    \alpha^2
    \sum_{i=2}^{N-1} [\MarkovChainUStateI{i} - \MarkovChainUuStateI{i}]^2
    \\
    &\qquad
    +
    \alpha\,(1-\alpha)
    \sum_{i=2}^{N-2}
    2\,[\MarkovChainUStateI{i-1} - \MarkovChainUuStateI{i-1}]
    \,[\MarkovChainUStateI{i+1} - \MarkovChainUuStateI{i+1}]
    -
    \sum_{i=1}^{N-1}
    [\MarkovChainUStateI{i} - \MarkovChainUuStateI{i}]^2
  \end{align*}
  where we made use of the convention
  $
  \MarkovChainUStateI{0} \equiv
  \MarkovChainUStateI{N} \equiv
  \MarkovChainUuStateI{0} \equiv
  \MarkovChainUuStateI{N} \equiv
  0$.

  Note now that the assumption $\int P(d\alpha)\,\alpha = \frac{1}{2}$
  implies
  \begin{align*}
    \int P(d\alpha)\,\alpha^2
    &=
    \int P(d\alpha)\,(1-\alpha)^2
    =
    \varP + \frac{1}{4}
    \;,\qquad
    \int P(d\alpha)\,\alpha\,(1-\alpha)
    =
    \frac{1}{4}
    -
    \varP
  \end{align*}
  and hence
  \begin{align*}
    \frac{1}{\Lambda}
    \,\MarkovGeneratorUU[\distU{ \MarkovChainUState }{ \MarkovChainUuState }^2]
    &=
    \int P(d\alpha) (1-\alpha)^2
    \sum_{i=1}^{N-2} [\MarkovChainUStateI{i} - \MarkovChainUuStateI{i}]^2
    +
    \int P(d\alpha) \alpha^2
    \sum_{i=2}^{N-1} [\MarkovChainUStateI{i} - \MarkovChainUuStateI{i}]^2
    \\
    &\qquad
    +
    \int P(d\alpha) \alpha\,(1-\alpha)
    \sum_{i=2}^{N-2}
    2\,[\MarkovChainUStateI{i-1} - \MarkovChainUuStateI{i-1}]
    \,[\MarkovChainUStateI{i+1} - \MarkovChainUuStateI{i+1}]
    \\
    &\qquad
    -
    \sum_{i=1}^{N-1}
    [\MarkovChainUStateI{i} - \MarkovChainUuStateI{i}]^2
    \\
    &=
    -
    \frac{1-4\,\varP}{4}\,\Big[
    \sum_{i=1}^{N-1} 2\,[\MarkovChainUStateI{i} - \MarkovChainUuStateI{i}]^2
    -
    \sum_{i=2}^{N-2}
    2\,[\MarkovChainUStateI{i-1} - \MarkovChainUuStateI{i-1}]
    \,[\MarkovChainUStateI{i+1} - \MarkovChainUuStateI{i+1}]
    \Big]
    \\
    &\qquad
    -
    \frac{1 + 4\,\varP}{4}\,\Big[
    [\MarkovChainUStateI{1} - \MarkovChainUuStateI{1}]^2
    +
    [\MarkovChainUStateI{N-1} - \MarkovChainUuStateI{N-1}]^2
    \Big]
    \;.
  \end{align*}
  It is now straightforward to verify that
  \begin{align*}
    \MarkovGeneratorUU[\distU{ \MarkovChainUState }{ \MarkovChainUuState }^2]
    &=
    -
    \Lambda\,\frac{1 - 4\,\varP}{4}
    \,[\MarkovChainUState - \MarkovChainUuState]^T
    \,\contractionMatrixN{N-1}
    \,[\MarkovChainUState - \MarkovChainUuState]
    \\
    &\qquad
    -
    \Lambda\,\frac{1+4\,\varP}{4}
    \,\Big[
    [\MarkovChainUStateI{1} - \MarkovChainUuStateI{1}]^2
    +
    [\MarkovChainUStateI{N-1} - \MarkovChainUuStateI{N-1}]^2
    \Big]
    \;,
  \end{align*}
  where the matrix $\contractionMatrixN{N-1}$ was defined
  in \eqref{eqn_definition_contractionMatrix} above.

  Observe that by
  \Lemref{lem_spectrum_contractionMatrix}
  the smallest eigenvalue of
  $\contractionMatrixN{N-1}$
  equals
  $4\,\sin^2[ \frac{\pi}{N+1} ]$ if $N$ is odd,
  and
  $4\,\sin^2[ \frac{\pi}{N+2} ]$ if $N$ is even.
  Therefore,
  \begin{align*}
    \MarkovGeneratorUU[\distU{ \MarkovChainUState }{ \MarkovChainUuState }^2]
    &\leq
    -
    \Lambda\,\frac{1-4\,\varP}{4}
    \,[\MarkovChainUState - \MarkovChainUuState]^T
    \,\contractionMatrixN{N-1}
    \,[\MarkovChainUState - \MarkovChainUuState]
    \\
    &\leq
    -
    \Lambda\,[1 - 4\,\varP]
    \,\sin^2\Big[ \frac{\pi}{N+2} \Big]
    \,\distU{ \MarkovChainUState }{ \MarkovChainUuState }^2
  \end{align*}
  follows from the fact that
  $\contractionMatrixN{N-1}$ is a symmetric matrix, and
  $0 \leq 1 - 4\,\varP$.
\end{proof}

Let $\MarkovChainU$ and $\MarkovChainUu$ be any two random variables
on $\stateSpaceU_{\epsilon,N}$ with distribution
denoted by $\MarkovChainUdist$ and $\MarkovChainUudist$, respectively.
Recall that for $p \geq 1$ the Vaserstein-$p$ distance
is defined by
\begin{equation*}
  \VasersteinDistp{p}{\MarkovChainU}{\MarkovChainUu}
  \equiv
  \VasersteinDistp{p}{\MarkovChainUdist}{\MarkovChainUudist}
  =
  \inf_{\Gamma}
  \Big[
  \int_{ \stateSpaceU_{\epsilon,N} \times \stateSpaceU_{\epsilon,N} }
  \Gamma(d\MarkovChainUState, d\MarkovChainUuState)
  \,\distU{ \MarkovChainUState }{ \MarkovChainUuState }^p
  \Big]^{\frac{1}{p}}
  \;,
\end{equation*}
where the infimum is taken over all couplings $\Gamma$ of
$\MarkovChainUdist$ and $\MarkovChainUudist$. To shorten the notation we set
$\VasersteinDist{\MarkovChainUdist}{\MarkovChainUudist}
\equiv \VasersteinDistp{1}{\MarkovChainUdist}{\MarkovChainUudist}$
in the special case $p=1$.

\begin{proposition}[Rate of convergence in Vaserstein-$2$ distance]
  \label{prop_convergence_vaserstein}
  Assume that the transition kernel $P$ satisfies
  $\int P(d\alpha)\,\alpha = \frac{1}{2}$.
  Let $\MarkovChainUT{t}$ and $\MarkovChainUuT{t}$ be two
  Markov chains generated by $\MarkovGeneratorU$ on
  $\stateSpaceU_{\epsilon,N}$.
  Then for all $t \geq 0$
  \begin{align*}
    \VasersteinDistp{2}{\MarkovChainUT{t}}{\MarkovChainUuT{t}}
    &\leq
    \VasersteinDistp{2}{\MarkovChainUT{0}}{\MarkovChainUuT{0}}
    \,\exp\Big(
    -
    \frac{1}{2}\,\Lambda\,[1 - 4\,\varP ]
    \,\sin^2\Big[ \frac{\pi}{N+2} \Big]
    \,t
    \Big)
    \\
    &\leq
    \epsilon\,N\,\sqrt{N-1}
    \,\exp\Big(
    -
    \frac{1}{2}\,\Lambda\,[1 - 4\,\varP ]
    \,\sin^2\Big[ \frac{\pi}{N+2} \Big]
    \,t
    \Big)
    \;.
  \end{align*}
\end{proposition}
\begin{proof}
  Denote the distribution of the bivariate Markov process
  $(\MarkovChainUT{t}, \MarkovChainUuT{t})$
  with generator $\MarkovGeneratorUU$
  by $\Gamma_t(d\MarkovChainUState,d\MarkovChainUuState)$,
  and denote by $\MarkovChainUdistT{t}(d\MarkovChainUState)$
  and $\MarkovChainUudistT{t}(d\MarkovChainUuState)$
  the two marginals.

  Observe that the generator $\MarkovGeneratorUU$ of this bivariate process
  $(\MarkovChainUT{t}, \MarkovChainUuT{t})$
  is constructed in such a way that
  $\MarkovChainUT{t}$ and $\MarkovChainUuT{t}$ are Markov chains
  with generator $\MarkovGeneratorU$ whose distributions
  are given by
  $\MarkovChainUdistT{t}(d\MarkovChainUState)$
  and
  $\MarkovChainUudistT{t}(d\MarkovChainUuState)$,
  respectively.

  Therefore,
  $\Gamma_t(d\MarkovChainUState,d\MarkovChainUuState)$
  is a coupling of the two distributions
  $\MarkovChainUdistT{t}(d\MarkovChainUState)$
  and
  $\MarkovChainUudistT{t}(d\MarkovChainUuState)$
  for all $t\geq 0$.
  In particular,
  \begin{align*}
    \VasersteinDistp{2}{\MarkovChainUT{t}}{\MarkovChainUuT{t}}^2
    \leq
    \int_{\stateSpaceU_{\epsilon,N} \times \stateSpaceU_{\epsilon,N}}
    \Gamma_t(d\MarkovChainUState,d\MarkovChainUuState)
    \,\distU{ \MarkovChainUState }{ \MarkovChainUuState }^2
  \end{align*}
  follows from the very definition of the Vaserstein distance.

  By the Markov property of the bivariate chain
  \begin{align*}
    \distU{ \MarkovChainUT{t} }{ \MarkovChainUuT{t} }^2
    -
    \distU{ \MarkovChainUT{0} }{ \MarkovChainUuT{0} }^2
    -
    \int_0^t
    \MarkovGeneratorUU \distU{ \MarkovChainUT{s} }{ \MarkovChainUuT{s} }^2
    \,ds
  \end{align*}
  is a centered martingale. Hence for all $t\geq 0$
  \begin{align*}
    \expect \distU{ \MarkovChainUT{t} }{ \MarkovChainUuT{t} }^2
    =
    \expect \distU{ \MarkovChainUT{0} }{ \MarkovChainUuT{0} }^2
    +
    \int_0^t
    \expect
    \MarkovGeneratorUU \distU{ \MarkovChainUT{s} }{ \MarkovChainUuT{s} }^2
    \,ds
    \;.
  \end{align*}
  Differentiating with respect to $t$ and
  applying the estimate of
  \Propref{prop_average_contraction}
  yields
  \begin{align*}
    \frac{d}{dt}
    \expect[ \distU{ \MarkovChainUT{t} }{ \MarkovChainUuT{t} }^2 ]
    &\leq
    -
    \Lambda\,[1 - 4\,\varP ]
    \,\sin^2\Big[ \frac{\pi}{N+2} \Big]
    \,\expect[ \distU{ \MarkovChainUT{t} }{ \MarkovChainUuT{t} }^2 ]
    \;.
  \end{align*}
  Gronwall's inequality shows that
  \begin{align*}
    \VasersteinDistp{2}{\MarkovChainUT{t}}{\MarkovChainUuT{t}}^2
    &\leq
    \expect[ \distU{ \MarkovChainUT{t} }{ \MarkovChainUuT{t} }^2 ]
    \\
    &\leq
    \exp\Big(
    -
    \Lambda\,[1 - 4\,\varP ]
    \,\sin^2\Big[ \frac{\pi}{N+2} \Big]
    \,t
    \Big)
    \expect[ \distU{ \MarkovChainUT{0} }{ \MarkovChainUuT{0} }^2 ]
  \end{align*}
  for any initial distribution $\Gamma_0$ of the bivariate chain.

  Taking in the infimum over all couplings $\Gamma_0$ of
  $\MarkovChainUdistT{0}$ and $\MarkovChainUudistT{0}$ yields
  \begin{align*}
    \VasersteinDistp{2}{\MarkovChainUT{t}}{\MarkovChainUuT{t}}^2
    &\leq
    \exp\Big(
    -
    \Lambda\,[1 - 4\,\varP ]
    \,\sin^2\Big[ \frac{\pi}{N+2} \Big]
    \,t
    \Big)
    \,\VasersteinDistp{2}{\MarkovChainUT{0}}{\MarkovChainUuT{0}}^2
    \\
    &\leq
    \epsilon^2\,N^2\,(N-1)
    \,\exp\Big(
    -
    \Lambda\,[1 - 4\,\varP ]
    \,\sin^2\Big[ \frac{\pi}{N+2} \Big]
    \,t
    \Big)
  \end{align*}
  where the second inequality is due to the estimate on
  the diameter of $\stateSpaceU_{\epsilon,N}$
  provided in \Lemref{lem_diameter_stateSpace}.
\end{proof}

By definition of the metric $\dist{.}{.}$ on $\stateSpace_{\epsilon,N}$
in terms of $\distU{.}{.}$ it follows immediately from
\Propref{prop_convergence_vaserstein}
that there is at most one stationary distribution for $\MarkovChainEnergyT{t}$
on each
$\stateSpace_{\epsilon,N}$, and that the rate of convergence in the
associated Vaserstein distance is the same as the rate of convergence
for $\MarkovChainUT{t}$.

Furthermore, by assumption the process $\MarkovChainEnergyT{t}$ on
$\stateSpace_{\epsilon,N}$ generated by $\MarkovGenerator$
is stochastically continuous. Hence the compactness
of $\stateSpace_{\epsilon,N}$
(in the topology induced by the chosen metric)
allows us to apply the Bogolyubov-Krylov argument to
show that there is at least one stationary distribution.
This proves the following \Thmref{thm_ergodicity}.

\begin{theorem}[Ergodicity and mixing rate of $\MarkovChainEnergyT{t}$
  on each $\stateSpace_{\epsilon,N}$.]
  \label{thm_ergodicity}
  If the transition kernel $P$ satisfies
  $\int P(d\alpha)\,\alpha = \frac{1}{2}$, and
  $\varP < \frac{1}{4}$, then there exists a unique
  stationary distribution $\stationaryDistEN{\epsilon}{N}$ on
  $\stateSpace_{\epsilon,N}$. Furthermore,
  \begin{align*}
    \VasersteinDistp{2}{\MarkovChainEnergyT{t}}{\stationaryDistEN{\epsilon}{N}}
    &\leq
    \VasersteinDistp{2}{\MarkovChainEnergyT{0}}{\stationaryDistEN{\epsilon}{N}}
    \,\exp\Big(
    -
    \frac{1}{2}\,\Lambda\,[1 - 4\,\varP ]
    \,\sin^2\Big[ \frac{\pi}{N+2} \Big]
    \,t
    \Big)
    \\
    &\leq
    \epsilon\,N\,\sqrt{N-1}
    \,\exp\Big(
    -
    \frac{1}{2}\,\Lambda\,[1 - 4\,\varP ]
    \,\sin^2\Big[ \frac{\pi}{N+2} \Big]
    \,t
    \Big)
  \end{align*}
  holds for all $t$, and any initial distribution of
  $\MarkovChainEnergyT{0}$ on $\stateSpace_{\epsilon,N}$.
\end{theorem}

\subsection{$\LpMu{2}{\stationaryDistEN{\epsilon}{N}}$--Spectral gap}

In order to analyse the spectrum
of $\MarkovGenerator$ in $\LpMu{2}{\stationaryDistEN{\epsilon}{N}}$
we will make an extra assumption on the invariant measure
$\stationaryDistEN{\epsilon}{N}$. Recall that a measure $\EnergyDist$ is
called reversible under $\MarkovGenerator$ if
for all bounded
$f \from \stateSpace_{\epsilon,N} \times \stateSpace_{\epsilon,N} \to \reals$
\begin{equation}
  \label{eqn_definitionReversibility}
  \int \MarkovChainUdist(d\EnergyState)
  \,[\MarkovGenerator f(., \EnergyState)](\EnergyState)
  =
  \int \MarkovChainUdist(d\EnergyState)
  \,[\MarkovGenerator f(\EnergyState, .)](\EnergyState)
\end{equation}
holds.
In particular, considering functions $f$ of the form
$ f(\EnergyState, \EEnergyState) = F(\EnergyState) $
for some bounded $F \from \stateSpace_{\epsilon,N} \to \reals$ shows that
$\MarkovChainUdist$ must be invariant under $\MarkovGenerator$.

Furthermore, $\MarkovGenerator$
acts on $\LpMu{2}{\MarkovChainUdist}$
as a bounded, self-adjoint negative semi-definite operator.
An estimate on the size of its spectral gap is provided in
\Thmref{thm_spectrumL2_reversible_constantRates} below.
Because the result of the following \Lemref{lem_generalSpectralProperty}
will play a central role in the proof of
\Thmref{thm_spectrumL2_reversible_constantRates}
we include the details of this well-known result for completeness.

\begin{lemma}[Auxiliary estimate on the spectrum of a self-adjoint operator]
  \label{lem_generalSpectralProperty}
  Let $H$ be a real (or complex)
  Hilbert space and $T \from H \to H$ a bounded, self-adjoint
  linear operator. Suppose there exists a constant
  $0 \leq \gamma$
  and a dense subspace $G \subset H$ on which
  for all $g\in G$ and $f\in H$ there exists a constant $C_{f,g}>0$
  such that
  $| \scalar{f}{T^n g}| \leq C_{f,g}\,\gamma^n$ for all $n\geq 1$.
  Then the spectrum of $T$ is contained in
  $[-\gamma,\gamma]$.
\end{lemma}
\begin{proof}
  The classical spectral theory of bounded self-adjoint linear operators
  \cite{MR1070713} states that
  the spectrum $\spectrum{T}$ of $T$ is a compact interval in
  $[-\norm{T}, \norm{T}]$,
  and there exists a unique spectral measure $\spectrumE{d\lambda}$ such that
  for any $f,g\in H$
  \begin{align*}
    1
    =
    \int_\reals \spectrumE{d\lambda}
    \;,\qquad
    T^n = \int_\reals \lambda^n\,\spectrumE{d\lambda}
    \;,\qquad
    \scalar{T^n f}{g}
    =
    \int_\reals \lambda^n\,\scalar{\spectrumE{d\lambda} f}{g}
  \end{align*}
  where $\spectrumE{d\lambda}$ is supported on $\spectrum{T}$,
  and $m_{f,g}(d\lambda) \equiv \scalar{\spectrumE{d\lambda} f}{g}$
  is a finite signed measure on $\spectrum{T}$, whose total variation
  norm satisfies
  $\TVnorm{m_{f,g}} \leq \norm{f}\,\norm{g}$.

  Suppose that the spectrum $\spectrum{T}$ of $T$ is not
  contained in $[-\gamma,\gamma]$. Then there exists $s>\gamma$
  such that for
  $
  S_s = (-\infty,-s) \cup (s,\infty)
  $
  the projection
  $\spectrumE{ S_s }$
  is nonzero. Hence there exists a nonzero $f_s \in H$ with
  $\spectrumE{ S_s } f_s = f_s$.
  In particular,
  \begin{align*}
    \norm{f_s}^2
    &=
    \int_{\spectrum{T}} m_{f_s, f_s}(d\lambda)
    =
    \int_{ S_s } m_{f_s, f_s}(d\lambda)
    >0
    \;,
  \end{align*}
  because the support of the measure $m_{f_s, f_s}(d\lambda)$
  is contained in $S_s$ by choice of $f_s$.
  In particular, $m_{f_s,f_s} \neq 0$.

  For any $g\in G$, and all $n\geq 0$ we have
  \begin{align*}
    \frac{1}{\gamma^{2\,n}} \scalar{f_s}{T^{2\,n} g}
    &=
    \frac{1}{\gamma^{2\,n}} \scalar{T^{2\,n} f_s}{g}
    =
    \int_{ S_s }
    \Big| \frac{\lambda}{\gamma} \Big|^{2\,n}
    \,m_{f_s,g}(d\lambda)
    \;.
  \end{align*}
  Due to the assumption on $G$ we also have that
  \begin{align*}
    \Big|
    \frac{1}{\gamma^{2\,n}} \scalar{f_s}{T^{2\,n} g}
    \Big|
    \leq
    C_{f_s,g}
  \end{align*}
  Since $m_{f_s,g}$ is a finite measure, and
  $|\frac{\lambda}{\gamma}| \geq \frac{s}{\gamma} > 1$ on
  its support, the boundedness of the above expression for all $n$
  can only be satisfied if in fact $m_{f_s,g} = 0$.

  Thus we have shown that $m_{f_s,f_s} \neq 0$, but $m_{f_s, g}=0$ for
  all $g \in G$. Since $m_{f_s, g}$ is continuous in $g$ (in fact linear
  and bounded) the denseness of $G$ implies that there exists
  a sequence $(g_n)_{n\geq 1} \subset G$ such that $g_n \to f_s$ in $H$, and
  hence $0 = m_{f_s, g_n} \to m_{f_s, f_s} \neq 0$. This is a contradiction
  to continuity. Therefore the assumption on $s$ must have been wrong, so
  that for all $s>\gamma$ the projection
  $\spectrumE{ S_s }$ must be zero. And since $\lambda\in \reals$ is
  in the resolvent set of $T$ if and only if there exists an
  open neighborhood $S$ of $\lambda$ such that $\spectrumE{S}=0$
  it follows that $\spectrum{T} \subset [-\gamma,\gamma]$.
\end{proof}

\begin{lemma}[Lipschitz contraction]
  \label{lem_LipschitzContraction}
  Let $A \from \stateSpace_{\epsilon,N} \to \reals$ be a Lipschitz
  continuous function with respect to the distance $\dist{.}{.}$,
  and set
  $
  A_t(\EnergyState)
  =
  \expect[ A( \MarkovChainEnergyT{t})
  \given \MarkovChainEnergyT{t} = \EnergyState]
  $
  for all $t \geq 0$
  and $\EnergyState \in \stateSpace_{\epsilon,N}$.
  Then $A_t$ is Lipschitz continuous with Lipschitz constant
  \begin{align*}
    \Lip{A_t}
    \leq
    \Lip{A}\,
    \exp\Big(
    -
    \frac{1}{2}\,\Lambda\,[1 - 4\,\varP ]
    \,\sin^2\Big[ \frac{\pi}{N+2} \Big]
    \,t
    \Big)
  \end{align*}
  for all $t\geq 0$.
\end{lemma}
\begin{proof}
  By Jensen's inequality it follows immediately from
  the very definition of the Vaserstein distance that
  $
  \VasersteinDistp{p_1}{\MarkovChainEnergyT{t}}{\MarkovChainEEnergyT{t}}
  \leq
  \VasersteinDistp{p_2}{\MarkovChainEnergyT{t}}{\MarkovChainEEnergyT{t}}
  $
  for all $1 \leq p_1 \leq p_2$. Therefore
  it follows from \Propref{prop_convergence_vaserstein} that
  \begin{align*}
    \VasersteinDistp{1}{\MarkovChainEnergyT{t}}{\MarkovChainEEnergyT{t}}
    &\leq
    \VasersteinDistp{2}{\MarkovChainEnergyT{0}}{\MarkovChainEEnergyT{0}}
    \,\exp\Big(
    -
    \frac{1}{2}\,\Lambda\,[1 - 4\,\varP ]
    \,\sin^2\Big[ \frac{\pi}{N+2} \Big]
    \,t
    \Big)
  \end{align*}
  for any joint distribution of
  $(\MarkovChainEnergyT{0}, \MarkovChainEEnergyT{0})$
  on $\stateSpace_{\epsilon,N} \times \stateSpace_{\epsilon,N}$.

  Note that $\stateSpace_{\epsilon,N}$ is compact, and hence
  \begin{align*}
    \sup_{\Lip{A} \leq 1}
    |
    \expect A( \MarkovChainEnergyT{t})
    -
    \expect A(\MarkovChainEEnergyT{t})
    |
    &=
    \VasersteinDistp{1}{\MarkovChainEnergyT{t}}{\MarkovChainEEnergyT{t}}
  \end{align*}
  which is the well-know Kantorovich-Rubinstein
  duality theorem for the Vaserstein-$1$ metric.

  Using the specific initial distribution
  $(\MarkovChainEnergyT{0}, \MarkovChainEEnergyT{0})
  =
  (\EnergyState, \EEnergyState)$
  on $\stateSpace_{\epsilon,N} \times \stateSpace_{\epsilon,N}$
  we obtain
  \begin{align*}
    |A_t(\EnergyState) - A_t(\EEnergyState)|
    &\leq
    \Lip{A}\,
    \VasersteinDistp{1}{\MarkovChainEnergyT{t}}{\MarkovChainEEnergyT{t}}
    \\
    &\leq
    \Lip{A}\,
    \dist{\EnergyState}{\EEnergyState}
    \,\exp\Big(
    -
    \frac{1}{2}\,\Lambda\,[1 - 4\,\varP ]
    \,\sin^2\Big[ \frac{\pi}{N+2} \Big]
    \,t
    \Big)
  \end{align*}
  because in this case
  $
  \VasersteinDistp{2}{\MarkovChainEnergyT{0}}{\MarkovChainEEnergyT{0}}
  =
  \dist{\EnergyState}{\EEnergyState}
  $.
  And since
  $
  \EnergyState,\EEnergyState
  \in\stateSpace_{\epsilon,N}
  $
  are arbitrary we see that $A_t$ is Lipschitz continuous with
  the claimed estimate on its Lipschitz constant.
\end{proof}

Combining now the result of \Lemref{lem_LipschitzContraction}
with that of \Lemref{lem_generalSpectralProperty} we
are in a position to estimate the spectral gap of $\MarkovGenerator$
acting on
$\LpMu{2}{\stationaryDistEN{\epsilon}{N}}$, provided
we assume that the stationary distribution
$\stationaryDistEN{\epsilon}{N}$ is reversible.
In this case $\MarkovGenerator$ is
a self-adjoint, bounded, negative semi-definite operator
on $\LpMu{2}{\stationaryDistEN{\epsilon}{N}}$.

\begin{theorem}[$\LpMu{2}{\stationaryDistEN{\epsilon}{N}}$--spectral gap for
  reversible $\stationaryDistEN{\epsilon}{N}$]
  \label{thm_spectrumL2_reversible_constantRates}
  Suppose that $P$ satisfies $\int P(d\alpha)\,\alpha = \frac{1}{2}$
  and $\varP < \frac{1}{4}$. If the stationary distribution
  $\stationaryDistEN{\epsilon}{N}$ of $\MarkovChainEnergyT{t}$
  on $\stateSpace_{\epsilon,N}$ is reversible, then
  \begin{align*}
    \spectrum{\MarkovGenerator}
    \subset
    \Big(
    -\infty,
    -\frac{1}{2}\,\Lambda\,[1 - 4\,\varP ]
    \,\sin^2\Big[ \frac{\pi}{N+2} \Big]
    \Big]
    \cup
    \{0\}
    \;,
  \end{align*}
  where $0$ is a simple eigenvalue corresponding to the constant eigenfunction.
\end{theorem}
\begin{proof}
  By assumption $\MarkovGenerator$ generates
  a self-adjoint, positive semi-definite contraction semigroup
  $e^{t \,\MarkovGenerator}$ on $\LpMu{2}{\stationaryDistEN{\epsilon}{N}}$,
  which satisfies $e^{t\,\MarkovGenerator} 1 = 1$.
  Therefore, the subspace $H$ of
  $\LpMu{2}{\stationaryDistEN{\epsilon}{N}}$
  consisting of functions perpendicular to the constant functions
  is invariant. Hence, the decomposition
  $\LpMu{2}{\stationaryDistEN{\epsilon}{N}} = H \oplus \Span\{1\}$
  is invariant under $e^{t\,\MarkovGenerator}$, and
  $e^{t\,\MarkovGenerator}$ may be restricted to $H$.

  Furthermore, it is a consequence of Lusin's theorem
  \cite{MR924157}
  that the set of
  Lipschitz continuous functions on $\stateSpace_{\epsilon,N}$
  is dense in $\LpMu{2}{\stationaryDistEN{\epsilon}{N}}$. Hence
  the set $G$ of Lipschitz continuous functions $A$
  on $\stateSpace_{\epsilon,N}$ with
  $\int \stationaryDistEN{\epsilon}{N}(d\EnergyState)
  \,A(\EnergyState) =0$ is dense in $H$.

  By \Lemref{lem_diameter_stateSpace} and the mean value theorem,
  for any $f\in H$ and $g \in G$
  \begin{align*}
    |\scalar{f}{g}|
    &\leq
    \norm{f}\,\norm{g}
    \leq
    \norm{f}\,\diam{\stateSpace_{\epsilon,N}}
    \,\Lip{g}
    \leq
    \norm{f}\,\epsilon\,N\,\sqrt{N-1}\,\Lip{g}
  \end{align*}
  and hence
  \begin{align*}
    |\scalar{f}{e^{n\,t\,\MarkovGenerator} g}|
    &\leq
    \norm{f}\,\epsilon\,N\,\sqrt{N-1}\,\Lip{g}
    \exp\Big(
    -
    \frac{1}{2}\,\Lambda\,[1 - 4\,\varP ]
    \,\sin^2\Big[ \frac{\pi}{N+2} \Big]
    \,t
    \Big)^n
  \end{align*}
  follows from \Lemref{lem_LipschitzContraction} for all $n\geq 0$.

  Since $e^{t\,\MarkovGenerator}$ is a positive operator
  the result of \Lemref{lem_generalSpectralProperty}
  yields
  \begin{align*}
    \spectrum{e^{t\,\MarkovGenerator} |_{H}}
    \subset
    \Big(
    0,
    \exp\Big(
    -
    \frac{1}{2}\,\Lambda\,[1 - 4\,\varP ]
    \,\sin^2\Big[ \frac{\pi}{N+2} \Big]
    \,t
    \Big)
    \Big]
    \;.
  \end{align*}
  This implies
  \begin{align*}
    \spectrum{\MarkovGenerator|_H}
    =
    \frac{1}{t}\,\log \spectrum{ e^{t \,\MarkovGenerator} }
    \subset
    \Big(
    -
    \infty,
    -
    \frac{1}{2}\,\Lambda\,[1 - 4\,\varP ]
    \,\sin^2\Big[ \frac{\pi}{N+2} \Big]
    \Big]
    \;,
  \end{align*}
  which finishes the proof.
\end{proof}

\begin{remark}
  From the proof of
  \Thmref{thm_spectrumL2_reversible_constantRates}
  it is clear that the abstract result
  \Lemref{lem_generalSpectralProperty}
  shows that an estimate on the exponential rate of weak convergence
  of $\MarkovChainEnergyT{t}$
  in Vaserstein-$1$ distance automatically yields an estimate on
  the spectral gap of $\MarkovGenerator$ on $\LpMu{2}{\stationaryDist}$,
  provided that the stationary distribution $\stationaryDist$
  is reversible.
  And since convergence in Vaserstein-$1$ distance can be
  controlled by two different approaches (recall the Kantorovich-Rubinstein
  duality theorem) we expect this general result to be also useful
  in other settings to prove estimates on $\Lp{2}$ spectral gaps.
\end{remark}

\begin{remark}
  All results of this section are essentially consequences of
  \Propref{prop_average_contraction}
  and
  \Lemref{lem_generalSpectralProperty}.
  And since the statement of
  \Propref{prop_average_contraction}
  is readily rephrased for the embedded discrete time
  Markov chain with transition operator
  \begin{align*}
    \transitionOperator A(\EnergyState)
    =
    A(\EnergyState)
    +
    \frac{1}{N-1}\,\frac{1}{\Lambda}\,
    \MarkovGenerator A(\EnergyState)
    =
    \sum_{i=1}^{N-1} \frac{1}{N-1}
    \int P(d\alpha)
    \,A( \mapTIA{i}{\alpha} \EnergyState)
  \end{align*}
  the results of this section all carry over (essentially verbatim) to the
  discrete time setting. One only has to multiply
  the rate of convergence (and hence the spectral gap)
  by $\frac{1}{N-1}\,\frac{1}{\Lambda}$ in the results for continuous time
  to obtain the corresponding results for the discrete time setting.
\end{remark}

\section{Spectral gap in $\LpMu{2}{\stationaryDistEN{\epsilon}{N}}$
for the general case}
\label{sect_gap_generalCase}

Now we consider the general situation
where the continuous-time Markov process $\MarkovChainEnergyT{t}$
is generated by the infinitesimal generator $\MarkovGenerator$ given in
\eqref{eqn_definitionGenerator}.
Suppose that $\stationaryDistEN{\epsilon}{N}$ is a reversible measure for
$\MarkovGenerator$. Then the associated Dirichlet form
\begin{subequations}
  \label{eqn_definitionDirichletForm}
  \begin{equation}
    \MarkovDirichletForm_{\epsilon,N}(A)
    =
    \int \stationaryDistEN{\epsilon}{N}(d\EnergyState)
    \,A(\EnergyState)
    \,[-\MarkovGenerator A](\EnergyState)
  \end{equation}
  is defined for all $A \in \LpMu{2}{\stationaryDistEN{\epsilon}{N}}$, and
  has the representation
  \begin{equation}
    \MarkovDirichletForm_{\epsilon,N}(A)
    =
    \frac{1}{2}
    \sum_{i=1}^{N-1}
    \int \stationaryDistEN{\epsilon}{N}(d\EnergyState)
    \,\rateFunction( \EnergyStateI{i}, \EnergyStateI{i+1} )
    \int \transitionP(\EnergyStateI{i}, \EnergyStateI{i+1} ,d\alpha )
    \,[A( \mapTIA{i}{\alpha} \EnergyState)
    -
    A(\EnergyState) ]^2
    \;.
  \end{equation}
\end{subequations}

The basic idea to prove convergence rates for $\MarkovChainEnergyT{t}$
is to compare the spectral gap of its generator $\MarkovGenerator$
to a suitably chosen reference process of the type
\eqref{eqn_definitionGenerator_reference}
considered in \Sectionref{section_reference}.
In order to distinguish these two generators we use
a superscript $\star$
\begin{align*}
  \MarkovGenerator^\star A(\EnergyState)
  &=
  \Lambda^\star
  \sum_{i=1}^{N-1}
  \int P^\star(d\alpha)
  \,[A( \mapTIA{i}{\alpha} \EnergyState)
  -
  A(\EnergyState) ]
  \\
  \MarkovDirichletForm^\star_{\epsilon,N}(A)
  &=
  \frac{1}{2}
  \int \stationaryDistEN{\epsilon}{N}^\star(d\EnergyState)
  \sum_{i=1}^{N-1}
  \Lambda^\star
  \int P^\star(d\alpha)
  \,[A( \mapTIA{i}{\alpha} \EnergyState)
  -
  A(\EnergyState) ]^2
\end{align*}
to denote the invariant measure, the generator and the corresponding
Dirichlet form of the reference process.

\begin{theorem}[Spectral gap for $\MarkovGenerator$]
  \label{thm_comparisionTheorem}
  Fix $\epsilon>0$ and $N$, and let $\stationaryDistEN{\epsilon}{N}$
  be a reversible stationary distribution of $\MarkovGenerator$
  on $\stateSpace_{\epsilon,N}$.
  Suppose that there exist a constant $\Lambda^\star>0$
  and a probability measure $P^\star$ on $[0,1]$ with mean
  $\int P^\star(d\alpha)\,\alpha = \frac{1}{2}$ and
  variance $\varPstar < \frac{1}{4}$
  such that the following are satisfied:
  \begin{enumerate}[(i)]
    \item
      \label{condition_lowerBound_rate}
      The rate function $\rateFunction$
      satisfies
      $
      \rateFunction(\EnergyStateI{i}, \EnergyStateI{i+1})
      \geq \Lambda^\star
      $
      for $\stationaryDistEN{\epsilon}{N}$--almost all
      $\EnergyState \in\stateSpace_{\epsilon,N}$,
      and all $1 \leq i \leq N-1$.

    \item
      \label{condition_mixing_P}
      There exists a constant $\beta>0$
      such that $P$ satisfies the minorization condition
      $
      P( \EnergyStateI{i}, \EnergyStateI{i+1}, .)
      \geq
      \beta\,P^\star(.)
      $
      for $\stationaryDistEN{\epsilon}{N}$--almost all
      $\EnergyState \in\stateSpace_{\epsilon,N}$,
      and all $1 \leq i \leq N-1$.

    \item
      \label{condition_reversiblity}
      The unique
      (recall \Thmref{thm_ergodicity})
      stationary distribution
      $\stationaryDistEN{\epsilon}{N}^\star$ of
      $\MarkovGenerator^\star$ on $\stateSpace_{\epsilon,N}$
      (corresponding to $\Lambda^\star$ and $P^\star$)
      is reversible.

    \item
      \label{condition_comparison_dists}
      The measures
      $\stationaryDistEN{\epsilon}{N}$
      and
      $\stationaryDistEN{\epsilon}{N}^\star$
      are uniformly equivalent, i.e.
      there exist two constants
      $0 < C_{\epsilon}^{-}\leq C_{\epsilon}^{+} < \infty$
      such that their Radon-Nikodym derivative satisfies
      $
      C_{\epsilon}^{-}
      \leq
      \frac{
      \stationaryDistEN{\epsilon}{N}(d\EnergyState)
      }{
      \stationaryDistEN{\epsilon}{N}^\star(d\EnergyState)
      }
      \leq
      C_{\epsilon}^{+}
      $ for all $N$.
  \end{enumerate}
  Then the spectrum of $\MarkovGenerator$ in
  $\LpMu{2}{\stationaryDistEN{\epsilon}{N}}$ satisfies
  \begin{equation*}
    \spectrum{\MarkovGenerator}
    \subset
    \Big(-\infty,
    -
    \beta\,\frac{C_{\epsilon}^{-}}{C_{\epsilon}^{+}}
    \,\Lambda^\star
    \,\frac{1}{2}\,[1 - 4\,\varPstar ]
    \,\sin^2\Big[ \frac{\pi}{N+2} \Big]
    \Big]
    \cup \{0\}
    \;,
  \end{equation*}
  where $0$ is a simple eigenvalue.
\end{theorem}

\begin{remark}
  Later we will see that - apart from
  condition \eqref{condition_lowerBound_rate} - the conditions of
  \Thmref{thm_comparisionTheorem} are fulfilled in a wide range
  of models of mechanical origin, interesting to us.
  Indeed, in \Thmref{thm_reversibleProductMeasures}
  we will prove
  a characterization of reversible measures of a particular type.
  Among others, it will provide the existence of reversible
  stationary measures for a large class rate functions $\rateFunction$
  and transition kernels $P$.  This result, in particular, addresses
  conditions
  \eqref{condition_reversiblity}
  and
  \eqref{condition_comparison_dists}
  in the above
  \Thmref{thm_comparisionTheorem}
  in quite satisfactory generality.
  Also, in \Sectionref{section_examples}, we show that
  \eqref{condition_mixing_P} is satisfied, for instance, in the
  Gaspard-Gilbert model with three-dimensional balls.
  Finally, \eqref{condition_lowerBound_rate} is the consequence
  of our method. Nevertheless establishing hydrodynamical limit transition
  is a great challenge even under our conditions and for doing
  it our theorems serve as an excellent background. Finally we note
  that the applicability of the statement in its present form seems
  to be restricted to models where
  $\stationaryDistEN{\epsilon}{N} = \stationaryDistEN{\epsilon}{N}^\star$
  therefore the weakening of condition \eqref{condition_comparison_dists}
  would also be desirable.
\end{remark}

\begin{proof}
  Since we assume reversibility the generator is self-adjoint,
  and hence we have the
  following variational characterization
  \begin{align*}
    \gamma = \inf\Big\{
    \frac{ \MarkovDirichletForm_{\epsilon,N}(A) }{\var_{\epsilon,N}(A)}
    \with
    A \in \LpMu{2}{\stationaryDistEN{\epsilon}{N}}
    ,\;
    \var_{\epsilon,N}(A) \neq 0
    \Big\}
  \end{align*}
  of the spectral gap $\gamma$ of $\MarkovGenerator$
  acting on $\LpMu{2}{\stationaryDistEN{\epsilon}{N}}$, where
  $\var_{\epsilon,N}(A)$ denotes the variance of $A$ with respect
  to $\stationaryDistEN{\epsilon}{N}$.

  By assumption we can compare the measures
  $\stationaryDistEN{\epsilon}{N}$,
  $\stationaryDistEN{\epsilon}{N}^\star$, and
  $P$, $P^\star$,
  so that for the Dirichlet form, recall
  \eqref{eqn_definitionDirichletForm}, we obtain
  the estimate
  \begin{align*}
    \MarkovDirichletForm_{\epsilon,N}(A)
    &=
    \frac{1}{2}
    \sum_{i=1}^{N-1}
    \int \stationaryDistEN{\epsilon}{N}(d\EnergyState)
    \,\rateFunction( \EnergyStateI{i}, \EnergyStateI{i+1} )
    \int \transitionP(\EnergyStateI{i}, \EnergyStateI{i+1} ,d\alpha )
    \,[A( \mapTIA{i}{\alpha} \EnergyState)
    -
    A(\EnergyState) ]^2
    \\
    &\geq
    \beta\,C_{\epsilon}^{-}
    \frac{1}{2}
    \sum_{i=1}^{N-1}
    \int \stationaryDistEN{\epsilon}{N}^\star(d\EnergyState)
    \,\Lambda^\star
    \int P^\star(d\alpha)
    \,[A( \mapTIA{i}{\alpha} \EnergyState)
    -
    A(\EnergyState) ]^2
  \end{align*}
  which is nothing else but
  $
  \MarkovDirichletForm_{\epsilon,N}(A)
  \geq
  \beta\,C_{\epsilon}^{-}
  \,\MarkovDirichletForm_{\epsilon,N}^\star(A)
  $
  for all $A \in \LpMu{2}{\stationaryDistEN{\epsilon}{N}}$.

  Furthermore, the variational characterization of the
  variance yields the estimate
  \begin{align*}
    \var_{\epsilon,N}(A)
    &=
    \inf_{c\in\reals}
    \int \stationaryDistEN{\epsilon}{N}(d\EnergyState)
    \,[A(\EnergyState) - c]^2
    =
    \inf_{c\in\reals}
    \int \stationaryDistEN{\epsilon}{N}^\star(d\EnergyState)
    \,\frac{
    \stationaryDistEN{\epsilon}{N}(d\EnergyState)
    }{
    \stationaryDistEN{\epsilon}{N}^\star(d\EnergyState)
    }
    \,[A(\EnergyState) - c]^2
    \\
    &\leq
    C_{\epsilon}^{+}
    \inf_{c\in\reals}
    \int \stationaryDistEN{\epsilon}{N}^\star(d\EnergyState)
    \,[A(\EnergyState) - c]^2
    =
    C_{\epsilon}^{+}\, \var_{\epsilon,N}^\star(A)
  \end{align*}
  for all $A \in \LpMu{2}{\stationaryDistEN{\epsilon}{N}}$.

  Combining both of the above estimates shows
  \begin{align*}
    \frac{ \MarkovDirichletForm_{\epsilon,N}(A) }{\var_{\epsilon,N}(A)}
    \geq
    \beta\,\frac{C_{\epsilon}^{-}}{C_{\epsilon}^{+}}
    \,\frac{ \MarkovDirichletForm_{\epsilon,N}^\star(A)
    }{\var_{\epsilon,N}^\star(A)}
  \end{align*}
  for any
  $A \in \LpMu{2}{\stationaryDistEN{\epsilon}{N}}$
  with
  $\var_{\epsilon,N}(A) \neq 0$. In other words, the spectral gap
  $\gamma$ of $\MarkovGenerator$ admits the estimate
  \begin{align*}
    \gamma
    \geq
    \beta\,\frac{C_{\epsilon}^{-}}{C_{\epsilon}^{+}}
    \,\inf\Big\{
    \frac{ \MarkovDirichletForm_{\epsilon,N}^\star(A)
    }{\var_{\epsilon,N}^\star(A)}
    \with
    A \in \LpMu{2}{\stationaryDistEN{\epsilon}{N}}
    ,\;
    \var_{\epsilon,N}(A) \neq 0
    \Big\}
    \;.
  \end{align*}

  Finally, note that the assumed bounds
  $
  C_{\epsilon}^{-}
  \leq
  \frac{
  \stationaryDistEN{\epsilon}{N}(d\EnergyState)
  }{
  \stationaryDistEN{\epsilon}{N}^\star(d\EnergyState)
  }
  \leq C_{\epsilon}^{+}
  $
  imply that
  $\LpMu{2}{\stationaryDistEN{\epsilon}{N}}
  =
  \LpMu{2}{\stationaryDistEN{\epsilon}{N}^\star}$
  so that the above estimate can be rewritten as
  \begin{equation*}
    \gamma
    \geq
    \beta\,\frac{C_{\epsilon}^{-}}{C_{\epsilon}^{+}}
    \,\gamma^\star
  \end{equation*}
  where $\gamma^\star$ denotes the spectral gap of
  $\MarkovGenerator^\star$ in
  $\LpMu{2}{\stationaryDistEN{\epsilon}{N}^\star}$.

  Now recall that by
  \Thmref{thm_spectrumL2_reversible_constantRates}
  \begin{equation*}
    \gamma^\star
    \geq
    \frac{1}{2}\,\Lambda^\star\,[1 - 4\,\varPstar ]
    \,\sin^2\Big[ \frac{\pi}{N+2} \Big]
  \end{equation*}
  which in turn shows for the spectral gap $\gamma$ of
  $\MarkovGenerator$
  \begin{equation*}
    \gamma
    \geq
    \beta\,\frac{C_{\epsilon}^{-}}{C_{\epsilon}^{+}}
    \,\Lambda^\star
    \,\frac{1}{2}\,[1 - 4\,\varPstar ]
    \,\sin^2\Big[ \frac{\pi}{N+2} \Big]
    \;,
  \end{equation*}
  which finishes the proof.
\end{proof}

\section{Classification of reversible product measures}
\label{sect_classification_products}

In this section we will characterize reversible product measures
of $\MarkovChainEnergyT{t}$. It is worth recalling at this
point that for any $N$ fixed
the sets $\stateSpace_{\epsilon,N} \subset \reals_+^N$
are invariant for the process for any choice of $\epsilon>0$.
And since these are simplexes there are no (non-trivial)
product measures
$
\MarkovChainEnergydist(d\EnergyState)
=
\MarkovChainEnergySitedist(d\EnergyStateI{1})
\cdots
\MarkovChainEnergySitedist(d\EnergyStateI{N})
$
supported by a single $\stateSpace_{\epsilon,N}$.
However, conditioning an invariant product measure on all of
$\reals_+^N$ to any $\stateSpace_{\epsilon,N}$ yields
an invariant measure on $\stateSpace_{\epsilon,N}$.
Therefore, we will consider product measures on
all of $\reals_+^N$ (canonical measures) instead on the ergodic components
$\stateSpace_{\epsilon,N}$ (micro-canonical measures).
And since our main convergence result
\Thmref{thm_comparisionTheorem} is for reversible
invariant measures, we consider here only
reversible product measures.

The first step in classifying all of them
is provided by \Lemref{lem_reversibleProductMeasures_systemSize},
which says that it suffices to consider $N=2$.
\begin{lemma}[Reversible product measures and system size]
  \label{lem_reversibleProductMeasures_systemSize}
  Let $\MarkovChainEnergySitedist$ be a probability measure
  on $\reals_+$. Then the product (probability) measure
  $
  \MarkovChainEnergydist(d\EnergyState)
  =
  \MarkovChainEnergySitedist(d\EnergyStateI{1})
  \cdots
  \MarkovChainEnergySitedist(d\EnergyStateI{N})
  $
  on $\reals_+^N$
  is reversible for $\MarkovChainEnergyT{t}$ (with generator
  \eqref{eqn_definitionGenerator})
  for some $N$
  if and only if it is reversible for $N=2$.
\end{lemma}
\begin{proof}
  Let $A \from \reals_+^N \times \reals_+^N \to \reals$ be bounded.
  To shorten the notation we use
  $[\MarkovGenerator A(\EnergyState,.)]$ to denote the function
  obtained by the action of the generator $\MarkovGenerator$
  on second variable of the function $A(\EnergyState, \EEnergyState)$,
  while treating the first variable as a parameter.
  Further we use
  $[\MarkovGenerator A(\EnergyState,.)](\EEnergyState)$ to denote
  the evaluation of the function
  $[\MarkovGenerator A(\EnergyState,.)]$ at the point $\EEnergyState$.
  Correspondingly, in
  $[\MarkovGenerator A(., \EEnergyState)]$ the second variable is treated
  as a parameter.

  By definition \eqref{eqn_definitionGenerator}
  of the generator $\MarkovGenerator$ we have
  \begin{align*}
    \int_{\reals_+^N}
    \MarkovChainEnergydist(d\EnergyState)
    \,[\MarkovGenerator A(., \EnergyState)](\EnergyState)
    &=
    \sum_{i=1}^{N-1}
    \int_{\reals_+^N}
    \MarkovChainEnergySitedist(d\EnergyStateI{1})
    \cdots
    \MarkovChainEnergySitedist(d\EnergyStateI{N})
    \,\rateFunction( \EnergyStateI{i}, \EnergyStateI{i+1} )
    \cdot
    \\
    &\qquad\qquad\cdot
    \int \transitionP(\EnergyStateI{i}, \EnergyStateI{i+1}, d\alpha)
    \,[
    A( \mapTIA{i}{\alpha} \EnergyState, \EnergyState)
    -
    A(\EnergyState, \EnergyState)
    ]
    \\
    =
    \sum_{i=1}^{N-1}
    \int_{\reals_+^2}
    \MarkovChainEnergySitedist(d\EnergyStateI{i})
    &
    \,\MarkovChainEnergySitedist(d\EnergyStateI{i+1})
    \,\rateFunction( \EnergyStateI{i}, \EnergyStateI{i+1} )
    \int \transitionP(\EnergyStateI{i}, \EnergyStateI{i+1}, d\alpha)
    \cdot
    \\
    \cdot
    \Big[
    A_{i,i+1}(
    &
    \alpha\,[\EnergyStateI{i} + \EnergyStateI{i+1}]
    ,
    (1-\alpha)\,[\EnergyStateI{i} + \EnergyStateI{i+1}]
    ,\EnergyStateI{i}, \EnergyStateI{i+1}
    )
    \\
    &\qquad\qquad\qquad\qquad
    -
    A_{i,i+1}(
    \EnergyStateI{i}, \EnergyStateI{i+1}
    ,\EnergyStateI{i}, \EnergyStateI{i+1}
    )
    \Big]
    \;,
  \end{align*}
  where we used the short hand notation
  \begin{align*}
    A_{i,i+1}(
    \EnergyStateI{i}, \EnergyStateI{i+1}
    ,
    \EEnergyStateI{i}, \EEnergyStateI{i+1}
    )
    &=
    \int_{\reals_+^{N-2}}
    \MarkovChainEnergySitedist(d\EnergyStateI{1})
    \cdots
    \MarkovChainEnergySitedist(d\EnergyStateI{i-1})
    \,\MarkovChainEnergySitedist(d\EnergyStateI{i+2})
    \cdots
    \MarkovChainEnergySitedist(d\EnergyStateI{N})
    \,A( \EnergyState, z_i)
    \;,
    \\
    z_i
    &\equiv
    (
    \EnergyStateI{1}, \ldots, \EnergyStateI{i-1}
    ,
    \EEnergyStateI{i}, \EEnergyStateI{i+1}
    ,
    \EnergyStateI{i+2},\ldots, \EnergyStateI{N}
    )
    \;.
  \end{align*}
  Recall that reversibility means
  $
  \int_{\reals_+^N}
  \MarkovChainEnergydist(d\EnergyState)
  \,[\MarkovGenerator A(., \EnergyState)](\EnergyState)
  =
  \int_{\reals_+^N}
  \MarkovChainEnergydist(d\EnergyState)
  \,[\MarkovGenerator A(\EnergyState,.)](\EnergyState)
  $, so that reversibility holds if and only if
  \begin{align*}
    \sum_{i=1}^{N-1}
    &
    \int_{\reals_+^2}
    \MarkovChainEnergySitedist(d\EnergyStateI{i})
    \,\MarkovChainEnergySitedist(d\EnergyStateI{i+1})
    \,\rateFunction(\EnergyStateI{i}, \EnergyStateI{i+1})
    \int P(\EnergyStateI{i}, \EnergyStateI{i+1}, d\alpha)
    \cdot
    \\
    &\qquad\qquad
    \cdot
    A_{i,i+1}(
    \alpha\,[\EnergyStateI{i} + \EnergyStateI{i+1}]
    ,
    (1-\alpha)\,[\EnergyStateI{i} + \EnergyStateI{i+1}]
    ,\EnergyStateI{i}, \EnergyStateI{i+1}
    )
    \\
    &=
    \sum_{i=1}^{N-1}
    \int_{\reals_+^2}
    \MarkovChainEnergySitedist(d\EnergyStateI{i})
    \,\MarkovChainEnergySitedist(d\EnergyStateI{i+1})
    \,\rateFunction( \EnergyStateI{i}, \EnergyStateI{i+1} )
    \int P(\EnergyStateI{i}, \EnergyStateI{i+1}, d\alpha)
    \\
    &\qquad\qquad\qquad
    \cdot
    A_{i,i+1}(
    \EnergyStateI{i}, \EnergyStateI{i+1}
    ,
    \alpha\,[\EnergyStateI{i} + \EnergyStateI{i+1}]
    ,
    (1-\alpha)\,[\EnergyStateI{i} + \EnergyStateI{i+1}]
    )
  \end{align*}
  for any bounded $A \from \reals_+^N \times \reals_+^N \to \reals$.

  In the particular case where
  $
  A(\EnergyState,\EEnergyState)
  =
  \phi(\EnergyStateI{1} ,\EEnergyStateI{1})
  $
  for some bounded $\phi \from \reals_+ \times \reals_+ \to \reals$
  \begin{align*}
    A_{1,2}(
    \EnergyStateI{1} ,\EnergyStateI{2}
    ,
    \EEnergyStateI{1} ,\EEnergyStateI{2}
    )
    &=
    \phi(\EnergyStateI{1} ,\EEnergyStateI{1})
    \\
    A_{i,i+1}(
    \EnergyStateI{i} ,\EnergyStateI{i+1}
    ,
    \EEnergyStateI{i} ,\EEnergyStateI{i+1}
    )
    &=
    \int_{\reals_+}
    \MarkovChainEnergySitedist(d\EnergyStateI{1})
    \,\phi(\EnergyStateI{1}, \EnergyStateI{1})
    \equiv
    \const
  \end{align*}
  for all $i=2,\ldots,N-1$. Hence reversibility requires
  \begin{align*}
    \int_{\reals_+^2}
    &
    \MarkovChainEnergySitedist(d\EnergyStateI{1})
    \,\MarkovChainEnergySitedist(d\EnergyStateI{2})
    \,\rateFunction(\EnergyStateI{1}, \EnergyStateI{2})
    \int P(\EnergyStateI{1}, \EnergyStateI{2}, d\alpha)
    \,\phi(
    \alpha\,[\EnergyStateI{1} + \EnergyStateI{2}]
    ,\EnergyStateI{1}
    )
    \\
    &=
    \int_{\reals_+^2}
    \MarkovChainEnergySitedist(d\EnergyStateI{1})
    \,\MarkovChainEnergySitedist(d\EnergyStateI{2})
    \,\rateFunction(\EnergyStateI{1}, \EnergyStateI{2})
    \int P(\EnergyStateI{1}, \EnergyStateI{2}, d\alpha)
    \,\phi(
    \EnergyStateI{1}
    ,
    \alpha\,[\EnergyStateI{1} + \EnergyStateI{2}]
    )
    \;.
  \end{align*}
  Consider now
  $
  A(\EnergyState,\EEnergyState)
  =
  \psi(
  \EnergyStateI{1}, \EnergyStateI{2}
  ,
  \EEnergyStateI{1} ,\EEnergyStateI{2}
  )
  $
  for some bounded
  $\psi \from \reals_+^2 \times \reals_+^2 \to \reals$.
  Then
  \begin{align*}
    A_{1,2}(
    \EnergyStateI{1} ,\EnergyStateI{2}
    ,
    \EEnergyStateI{1} ,\EEnergyStateI{2}
    )
    &=
    \psi(
    \EnergyStateI{1}, \EnergyStateI{2}
    ,
    \EEnergyStateI{1} ,\EEnergyStateI{2}
    )
    \\
    A_{2,3}(
    \EnergyStateI{2} ,\EnergyStateI{3}
    ,
    \EEnergyStateI{2} ,\EEnergyStateI{3}
    )
    &=
    \int_{\reals_+}
    \MarkovChainEnergySitedist(d\EnergyStateI{1})
    \,\psi(
    \EnergyStateI{1}, \EnergyStateI{2}
    ,
    \EnergyStateI{1} ,\EEnergyStateI{2}
    )
    \equiv
    \hat{\psi}(
    \EnergyStateI{2}
    ,
    \EEnergyStateI{2}
    )
    \\
    A_{i,i+1}(
    \EnergyStateI{i} ,\EnergyStateI{i+1}
    ,
    \EEnergyStateI{i} ,\EEnergyStateI{i+1}
    )
    &=
    \int_{\reals_+^2}
    \MarkovChainEnergySitedist(d\EnergyStateI{1})
    \,\MarkovChainEnergySitedist(d\EnergyStateI{2})
    \,\psi(
    \EnergyStateI{1}, \EnergyStateI{2}
    ,
    \EnergyStateI{1} ,\EnergyStateI{2}
    )
    \equiv
    \const
  \end{align*}
  for all $i=3,\ldots,N-1$. Combining this with the previous
  special case (applied to $\phi = \hat{\psi}$)
  shows that reversibility requires
  \begin{align*}
    \int_{\reals_+^2}
    &
    \MarkovChainEnergySitedist(d\EnergyStateI{1})
    \,\MarkovChainEnergySitedist(d\EnergyStateI{2})
    \,\rateFunction(\EnergyStateI{1}, \EnergyStateI{2})
    \int P(\EnergyStateI{1}, \EnergyStateI{2}, d\alpha)
    \cdot
    \\
    &\qquad\qquad\cdot
    \psi(
    \alpha\,[\EnergyStateI{1} + \EnergyStateI{2}]
    ,
    (1-\alpha)\,[\EnergyStateI{1} + \EnergyStateI{2}]
    ,\EnergyStateI{1}, \EnergyStateI{2}
    )
    \\
    &=
    \int_{\reals_+^2}
    \MarkovChainEnergySitedist(d\EnergyStateI{1})
    \,\MarkovChainEnergySitedist(d\EnergyStateI{2})
    \,\rateFunction(\EnergyStateI{1}, \EnergyStateI{2})
    \int P(\EnergyStateI{1}, \EnergyStateI{2}, d\alpha)
    \cdot
    \\
    &\qquad\qquad\qquad\cdot
    \psi(
    \EnergyStateI{1}, \EnergyStateI{2}
    ,
    \alpha\,[\EnergyStateI{1} + \EnergyStateI{2}]
    ,
    (1-\alpha)\,[\EnergyStateI{1} + \EnergyStateI{2}]
    )
  \end{align*}
  for any bounded test function
  $\psi \from \reals_+^2 \times \reals_+^2 \to \reals$.
  And since this is also sufficient for reversibility, it follows
  that reversibility of the product measure
  holds if and only if the above equality holds for all $\psi$.

  Finally, observe that this last expression is precisely
  the reversibility condition for $N=2$, which finishes the proof.
\end{proof}

The final expression in the above proof actually shows
that reversibility of the product measure is equivalent
to a slightly stronger statement than the one stated in
\Lemref{lem_reversibleProductMeasures_systemSize}.
Namely, because both integrands agree at $(0,0)$ the reversibility
of the product measure is equivalent to
\begin{equation}
  \label{eqn_exclusionOrigin}
  \begin{split}
  \int_{\reals_+^2 \setminus\{(0,0)\}}
  &
  \MarkovChainEnergySitedist(d\EnergyStateI{1})
  \,\MarkovChainEnergySitedist(d\EnergyStateI{2})
  \,\rateFunction(\EnergyStateI{1}, \EnergyStateI{2})
  \int P(\EnergyStateI{1}, \EnergyStateI{2}, d\alpha)
  \cdot
  \\
  &\qquad\qquad\cdot
  \psi(
  \alpha\,[\EnergyStateI{1} + \EnergyStateI{2}]
  ,
  (1-\alpha)\,[\EnergyStateI{1} + \EnergyStateI{2}]
  ,\EnergyStateI{1}, \EnergyStateI{2}
  )
  \\
  &=
  \int_{\reals_+^2 \setminus\{(0,0)\}}
  \MarkovChainEnergySitedist(d\EnergyStateI{1})
  \,\MarkovChainEnergySitedist(d\EnergyStateI{2})
  \,\rateFunction(\EnergyStateI{1}, \EnergyStateI{2})
  \int P(\EnergyStateI{1}, \EnergyStateI{2}, d\alpha)
  \cdot
  \\
  &\qquad\qquad\qquad\cdot
  \psi(
  \EnergyStateI{1}, \EnergyStateI{2}
  ,
  \alpha\,[\EnergyStateI{1} + \EnergyStateI{2}]
  ,
  (1-\alpha)\,[\EnergyStateI{1} + \EnergyStateI{2}]
  )
  \;.
  \end{split}
\end{equation}
This simplification
is relevant, because so far we have not ruled out yet the possibility
of $\MarkovChainEnergySitedist$ having an atom at $0$.

For the further analysis we will need to assume that
the rate function $\rateFunction$ and the transition kernel $P$
are of the form
\begin{equation}
  \label{eqn_assumption_Lambda_P}
  \begin{split}
    \rateFunction(\EnergyStateI{i}, \EnergyStateI{i+1})
    &=
    \rateSum(\EnergyStateI{i} + \EnergyStateI{i+1})
    \,\rateRatio\Big(
    \frac{\EnergyStateI{i}}{ \EnergyStateI{i} + \EnergyStateI{i+1}}
    \Big)
    \\
    P(\EnergyStateI{i}, \EnergyStateI{i+1}, d\alpha)
    &=
    P\Big( \frac{\EnergyStateI{i}}{ \EnergyStateI{i} + \EnergyStateI{i+1}}
    , d\alpha
    \Big)
    \;.
  \end{split}
\end{equation}
Here the subscripts $s$ and $r$ stand for ``sum'' and ``ratio'',
respectively.
Note that
$\frac{\EnergyStateI{i}}{ \EnergyStateI{i} + \EnergyStateI{i+1}}$
makes sense everywhere on $\reals_+^2 \setminus \{(0,0)\}$,
and by the above this set is all that we need to consider.
In \Sectionref{section_examples} below
we will see that the representation
\eqref{eqn_assumption_Lambda_P}
naturally occurs
in models originating from mechanical systems.

We have already shown that in order to classify reversible
product measures for arbitrary $N$ it is enough to study
the case $N=2$. This, however, is still not a completely straightforward
problem, since the answer might depend on
the rate functions $\rateSum$ and $\rateRatio$.
The next \Corref{cor_reversibleProductMeasures_rates}
simplifies this issue.

\begin{corollary}[Reversible product measures and rate functions]
  \label{cor_reversibleProductMeasures_rates}
  If $\rateSum(\eta) > 0$ for all $0 < \eta < \infty$, then
  the process has a reversible stationary product measure
  $\MarkovChainEnergydist$
  (as in \Lemref{lem_reversibleProductMeasures_systemSize})
  if and only if
  \begin{align*}
    \int_{\reals_+^2 \setminus\{(0,0)\}}
    &
    \MarkovChainEnergySitedist(d\EnergyStateI{1})
    \,\MarkovChainEnergySitedist(d\EnergyStateI{2})
    \,\rateRatio\Big(
    \frac{\EnergyStateI{1}}{
    \EnergyStateI{1} + \EnergyStateI{2}}
    \Big)
    \int P\Big(
    \frac{\EnergyStateI{1}}{
    \EnergyStateI{1} + \EnergyStateI{2}}
    ,d\alpha \Big)
    \cdot
    \\
    &\qquad\qquad
    \cdot
    \eta\Big(
    \EnergyStateI{1} + \EnergyStateI{2}
    , \alpha
    ,
    \frac{\EnergyStateI{1}}{
    \EnergyStateI{1} + \EnergyStateI{2}}
    \Big)
    \\
    &=
    \int_{\reals_+^2 \setminus\{(0,0)\}}
    \MarkovChainEnergySitedist(d\EnergyStateI{1})
    \,\MarkovChainEnergySitedist(d\EnergyStateI{2})
    \,\rateRatio\Big(
    \frac{\EnergyStateI{1}}{
    \EnergyStateI{1} + \EnergyStateI{2}}
    \Big)
    \int P\Big(
    \frac{\EnergyStateI{1}}{
    \EnergyStateI{1} + \EnergyStateI{2}}
    ,d\alpha \Big)
    \cdot
    \\
    &\qquad\qquad\qquad
    \cdot
    \eta\Big(
    \EnergyStateI{1} + \EnergyStateI{2}
    ,
    \frac{\EnergyStateI{1}}{
    \EnergyStateI{1} + \EnergyStateI{2}}
    , \alpha
    \Big)
  \end{align*}
  holds for all bounded
  $\eta \from \reals_+\setminus\{0\} \times [0,1]^2 \to \reals$.
\end{corollary}
\begin{proof}
  By \eqref{eqn_exclusionOrigin}
  reversibility of the product measure
  is equivalent to
  \begin{align*}
    &
    \int_{\reals_+^2 \setminus \{(0,0)\}}
    \MarkovChainEnergySitedist(d\EnergyStateI{1})
    \,\MarkovChainEnergySitedist(d\EnergyStateI{2})
    \,\rateSum( \EnergyStateI{1} + \EnergyStateI{2} )
    \,\rateRatio\Big(
    \frac{\EnergyStateI{1}}{
    \EnergyStateI{1} + \EnergyStateI{2}}
    \Big)
    \int P\Big(
    \frac{\EnergyStateI{1}}{
    \EnergyStateI{1} + \EnergyStateI{2}}
    ,d\alpha \Big)
    \cdot
    \\
    &\qquad\qquad
    \cdot
    \psi(
    \alpha\,[\EnergyStateI{1} + \EnergyStateI{2}]
    ,
    (1-\alpha)\,[\EnergyStateI{1} + \EnergyStateI{2}]
    ,\EnergyStateI{1}, \EnergyStateI{2}
    )
    \\
    &\qquad
    =
    \int_{\reals_+^2 \setminus \{(0,0)\}}
    \MarkovChainEnergySitedist(d\EnergyStateI{1})
    \,\MarkovChainEnergySitedist(d\EnergyStateI{2})
    \,\rateSum( \EnergyStateI{1} + \EnergyStateI{2} )
    \,\rateRatio\Big(
    \frac{\EnergyStateI{1}}{
    \EnergyStateI{1} + \EnergyStateI{2}}
    \Big)
    \int P\Big(
    \frac{\EnergyStateI{1}}{
    \EnergyStateI{1} + \EnergyStateI{2}}
    ,d\alpha \Big)
    \cdot
    \\
    &\qquad\qquad\qquad
    \cdot
    \psi(
    \EnergyStateI{1}, \EnergyStateI{2}
    ,
    \alpha\,[\EnergyStateI{1} + \EnergyStateI{2}]
    ,
    (1-\alpha)\,[\EnergyStateI{1} + \EnergyStateI{2}]
    )
  \end{align*}
  for any (non-negative) test function
  $\psi \from \reals^2\setminus\{(0,0)\} \times \reals_+^2 \to \reals$.
  On $\reals_+^2\setminus \{(0,0)\}$ the change of coordinates
  $
  (\EnergyStateI{1}, \EnergyStateI{2})
  \mapsto
  (
  \EnergyStateI{1} + \EnergyStateI{2}
  ,
  \frac{\EnergyStateI{1}}{
  \EnergyStateI{1} + \EnergyStateI{2}}
  )
  $
  is one-to-one, hence any such function $\psi$ may be recast as
  \begin{align*}
    \psi(
    \EnergyStateI{1}, \EnergyStateI{2}
    ,
    \EEnergyStateI{1}, \EEnergyStateI{2}
    )
    \equiv
    \eta\Big(
    \EnergyStateI{1} + \EnergyStateI{2}
    ,
    \frac{\EnergyStateI{1}}{
    \EnergyStateI{1} + \EnergyStateI{2}}
    ,
    \EEnergyStateI{1} + \EEnergyStateI{2}
    ,
    \frac{\EEnergyStateI{1}}{
    \EEnergyStateI{1} + \EEnergyStateI{2}}
    \Big)
  \end{align*}
  for some function $\eta \from (\reals_+ \times [0,1])^2 \to \reals$.
  Therefore reversibility holds if and only if
  \begin{align*}
    &
    \int_{\reals_+^2 \setminus\{(0,0)\}}
    \MarkovChainEnergySitedist(d\EnergyStateI{1})
    \,\MarkovChainEnergySitedist(d\EnergyStateI{2})
    \,\rateSum( \EnergyStateI{1} + \EnergyStateI{2} )
    \,\rateRatio\Big(
    \frac{\EnergyStateI{1}}{
    \EnergyStateI{1} + \EnergyStateI{2}}
    \Big)
    \int P\Big(
    \frac{\EnergyStateI{1}}{
    \EnergyStateI{1} + \EnergyStateI{2}}
    ,d\alpha \Big)
    \cdot
    \\
    &\qquad\qquad\qquad
    \cdot
    \eta\Big(
    \EnergyStateI{1} + \EnergyStateI{2}
    , \alpha
    ,
    \frac{\EnergyStateI{1}}{
    \EnergyStateI{1} + \EnergyStateI{2}}
    \Big)
    \\
    &\qquad
    =
    \int_{\reals_+^2 \setminus\{(0,0)\}}
    \MarkovChainEnergySitedist(d\EnergyStateI{1})
    \,\MarkovChainEnergySitedist(d\EnergyStateI{2})
    \,\rateSum( \EnergyStateI{1} + \EnergyStateI{2} )
    \,\rateRatio\Big(
    \frac{\EnergyStateI{1}}{
    \EnergyStateI{1} + \EnergyStateI{2}}
    \Big)
    \int P\Big(
    \frac{\EnergyStateI{1}}{
    \EnergyStateI{1} + \EnergyStateI{2}}
    ,d\alpha \Big)
    \cdot
    \\
    &\qquad\qquad\qquad\qquad
    \cdot
    \eta\Big(
    \EnergyStateI{1} + \EnergyStateI{2}
    ,
    \frac{\EnergyStateI{1}}{
    \EnergyStateI{1} + \EnergyStateI{2}}
    , \alpha
    \Big)
  \end{align*}
  holds for all
  $\eta \from \reals_+\setminus\{0\} \times [0,1]^2 \to \reals$.

  And since
  $
  \EnergyStateI{1} + \EnergyStateI{2} >0
  $
  our assumption on $\rateSum$ implies that
  $\rateSum$ is strictly positive, and hence may as well
  be combined with $\eta$, because $\eta$ is arbitrary.
  This finishes the proof.
\end{proof}

With \Lemref{lem_reversibleProductMeasures_systemSize},
and \Corref{cor_reversibleProductMeasures_rates}
we are now in a position to classify all reversible product
measures, which is the content of the following
\Thmref{thm_reversibleProductMeasures}.
This classification relies on a well-known fact
\cite{MR0069408}
about Gamma distributions. Namely, suppose that
$\MarkovChainEnergyI{1}$ and $\MarkovChainEnergyI{2}$
are two non-constant, independent, positive random variables.
Then
$\MarkovChainEnergyI{1} + \MarkovChainEnergyI{2}$ and
$
\frac{\MarkovChainEnergyI{1}}{ \MarkovChainEnergyI{1} + \MarkovChainEnergyI{2}}
$
are independent if and only if
$\MarkovChainEnergyI{1}$ and $\MarkovChainEnergyI{2}$
are independent, identically Gamma-distributed random variables.

In the theorem below we use the following notation:
For $\epsilon >0$ we denote by $\delta(\epsilon, d\alpha)$
the Dirac measure concentrated at $\epsilon$.

\begin{theorem}[Reversible product measures]
  \label{thm_reversibleProductMeasures}
  Suppose that the Markov chain on $[0,1]$ with
  transition kernel $P(\beta, d\alpha)$
  has a unique invariant distribution, say $p(.)$.
  Let $N$ be arbitrary, and suppose further that
  $\rateSum$ is such that $\rateSum(\sigma) > 0$ for all $\sigma>0$,
  and $\rateRatio(\beta) > 0$ for all $0<\beta<1$.
  Then the product measure
  $
  \MarkovChainEnergydist(d\EnergyState)
  =
  \MarkovChainEnergySitedist(d\EnergyStateI{1})
  \cdots
  \MarkovChainEnergySitedist(d\EnergyStateI{N})
  $
  is reversible for $\MarkovChainEnergyT{t}$
  if and only if
  $p$ is a reversible measure for the Markov chain generated by
  $P$,
  and either of the following two holds:
  \begin{enumerate}
    \item
      There exists $\epsilon> 0$ and $d>0$ such that
      \begin{align*}
        \MarkovChainEnergySitedist(d\EnergyStateI{1})
        &=
        \frac{d\EnergyStateI{1}}{\epsilon}
        \,\Big[
        \frac{ \EnergyStateI{1} }{\epsilon}
        \Big]^{\frac{d}{2}-1}
        \,\frac{e^{- \frac{ \EnergyStateI{1} }{\epsilon}}
        }{ \Gamma(\frac{d}{2} ) }
        \\
        p(d\beta)
        &=
        d\beta
        \,[ \beta\,(1-\beta) ]^{\frac{d}{2}-1}
        \,\frac{ \Gamma(d ) }{ \Gamma(\frac{d}{2} )^2 }
        \,\rateRatio(\beta)\,\frac{1}{Z}
      \end{align*}
      where $Z$ is the normalizing constant.

    \item
      There exists $\epsilon>0$ such that
      $
      \MarkovChainEnergySitedist(d\EnergyStateI{1})
      =
      \delta(\epsilon,d\EnergyStateI{1})
      $,
      $
      p(d\alpha)
      =
      \delta(\frac{1}{2},d\alpha)
      $,
      and
      $
      P(\frac{1}{2}, d\alpha)
      =
      \delta(\frac{1}{2}, d\alpha)
      $.
  \end{enumerate}
\end{theorem}
\begin{proof}
  From \Lemref{lem_reversibleProductMeasures_systemSize}
  we know that it suffices to consider $N=2$, and
  \Corref{cor_reversibleProductMeasures_rates}
  shows - as it is also clear intuitively - that
  the choice of $\rateSum$ is irrelevant,
  and that we only need to consider the process on
  $\reals_+^2 \setminus\{(0,0)\}$.

  Using the change of variables
  $\sigma = \EnergyStateI{1} + \EnergyStateI{2}$
  ,
  $
  \beta = \frac{\EnergyStateI{1}}{
  \EnergyStateI{1} + \EnergyStateI{2}}
  $
  on $\reals_+^2 \setminus\{(0,0)\}$
  we can disintegrate the product measure
  $
  \MarkovChainEnergySitedist(d\EnergyStateI{1})
  \,\MarkovChainEnergySitedist(d\EnergyStateI{2})
  $
  such that for any (bounded) $\eta \from \reals_+^2 \to \reals$
  we have
  \begin{align*}
    \int_{\reals_+^2 \setminus\{(0,0)\}}
    \MarkovChainEnergySitedist(d\EnergyStateI{1})
    \,\MarkovChainEnergySitedist(d\EnergyStateI{2})
    \,\eta( \EnergyStateI{1}, \EnergyStateI{2})
    =
    \int_{\reals_+ \setminus\{0\}}
    \MarkovChainEnergySiteSumdist(d\sigma)
    \int_{[0,1]}
    \MarkovChainEnergySiteRatiodist(\sigma,d\beta)
    \,\eta( \beta\,\sigma, (1-\beta)\,\sigma)
  \end{align*}
  where
  $\MarkovChainEnergySiteSumdist(.)$ is the distribution of the sum
  $\EnergyStateI{1} + \EnergyStateI{2}$
  and
  $\MarkovChainEnergySiteRatiodist(\sigma,.)$ is the conditional distribution
  of the ratio
  $
  \frac{\EnergyStateI{1}}{
  \EnergyStateI{1} + \EnergyStateI{2}}
  $
  given that
  $\EnergyStateI{1} + \EnergyStateI{2}=\sigma$.

  Using this notation the condition
  for the reversibility of the product measure
  of \Corref{cor_reversibleProductMeasures_rates}
  takes on the form
  \begin{align*}
    \int
    &
    \MarkovChainEnergySiteSumdist(d\sigma)
    \int
    \MarkovChainEnergySiteRatiodist(\sigma,d\beta)
    \,\rateRatio(\beta)
    \int P(\beta,d\alpha)
    \,\eta(\sigma, \alpha, \beta)
    \\
    &=
    \int
    \MarkovChainEnergySiteSumdist(d\sigma)
    \int
    \MarkovChainEnergySiteRatiodist(\sigma,d\beta)
    \,\rateRatio(\beta)
    \int P(\beta, d\alpha)
    \,\eta(\sigma, \beta, \alpha)
    \;.
  \end{align*}
  This holds if and only if for
  $\MarkovChainEnergySiteSumdist$--almost every $\sigma$
  \begin{equation}
    \label{eqn_reversibility_s_beta}
    \begin{split}
      \int
      \MarkovChainEnergySiteRatiodist(\sigma,d\beta)
      &
      \,\rateRatio(\beta)
      \int P(\beta,d\alpha)
      \,\tilde{\eta}(\alpha, \beta)
      \\
      &=
      \int
      \MarkovChainEnergySiteRatiodist(\sigma,d\beta)
      \,\rateRatio(\beta)
      \int P(\beta, d\alpha)
      \,\tilde{\eta}(\beta, \alpha)
    \end{split}
  \end{equation}
  for all bounded $\tilde{\eta} \from [0,1]\times [0,1] \to \reals$.

  Suppose now that the product measure is reversible.
  The special choice $\eta(\alpha,\beta) = \psi(\alpha)$
  for some $\psi \from [0,1] \to \reals$ thus shows that
  \begin{align*}
    \int
    \MarkovChainEnergySiteRatiodist(\sigma,d\beta)\,\rateRatio(\beta)
    \int P(\beta,d\alpha)
    \,\psi(\alpha)
    =
    \int
    \MarkovChainEnergySiteRatiodist(\sigma,d\beta)\,\rateRatio(\beta)
    \,\psi(\beta)
  \end{align*}
  for all $\psi$. In other words, the (not normalized) non-negative
  measure
  $\MarkovChainEnergySiteRatiodist(\sigma,d\beta)\,\rateRatio(\beta)$
  must be invariant under $P$.
  And since by assumption $P$ has a unique invariant distribution,
  denote it by $p$,
  it thus follows that
  \begin{equation*}
    \frac{1}{Z}\,\MarkovChainEnergySiteRatiodist(\sigma,d\beta)
    \,\rateRatio(\beta)
    =
    p(d\beta)
    \;,\qquad
    Z = \int \MarkovChainEnergySiteRatiodist(\sigma,d\beta)\,\rateRatio(\beta)
  \end{equation*}
  for $\MarkovChainEnergySiteSumdist$--almost every $\sigma$, where
  $Z>0$ by assumption on $\rateRatio$.

  In particular,
  this means that the conditional distribution
  $\MarkovChainEnergySiteRatiodist(\sigma,.)$
  of the ratio
  $
  \frac{\EnergyStateI{1}}{
  \EnergyStateI{1} + \EnergyStateI{2}}
  $
  given that
  $\sigma = \EnergyStateI{1} + \EnergyStateI{2}$
  actually is the same for all values of $\sigma$.
  In other words the sum
  $\EnergyStateI{1} + \EnergyStateI{2}$
  and the ratio
  $
  \frac{\EnergyStateI{1}}{
  \EnergyStateI{1} + \EnergyStateI{2}}
  $
  are independent.
  And since also
  $\EnergyStateI{1}$ and $\EnergyStateI{2}$
  are independent (by assumption) we conclude
  \cite{MR0069408}
  that either
  $\MarkovChainEnergySitedist$ is a point mass, i.e.
  $\MarkovChainEnergySitedist(d\EnergyStateI{1})
  =
  \delta(\epsilon, d\EnergyStateI{1})
  $
  for some $\epsilon>0$, or
  $\MarkovChainEnergySitedist$ is a Gamma distribution, i.e.
  \begin{equation*}
    \MarkovChainEnergySitedist(d\EnergyStateI{1})
    =
    \frac{d\EnergyStateI{1}}{\epsilon}
    \,\Big[
    \frac{ \EnergyStateI{1} }{\epsilon}
    \Big]^{\frac{d}{2}-1}
    \,\frac{e^{- \frac{ \EnergyStateI{1} }{\epsilon}}
    }{ \Gamma(\frac{d}{2} ) }
    \qquad
    ( 0 < \EnergyStateI{1} < \infty )
  \end{equation*}
  for some $\epsilon>0$ and $d>0$

  In the former case it follows
  \begin{align*}
    \MarkovChainEnergySiteSumdist(d\sigma)
    &=
    \delta(2\,\epsilon, d\sigma)
    \;,\qquad
    p(d\beta)
    =
    \MarkovChainEnergySiteRatiodist(d\beta)
    =
    \delta(\frac{1}{2},d\beta)
  \end{align*}
  for
  $\MarkovChainEnergySiteSumdist$,
  $\MarkovChainEnergySiteRatiodist$. Hence
  the reversibility condition \eqref{eqn_reversibility_s_beta}
  becomes
  $
  \int P(\frac{1}{2},d\alpha) \,\eta(\alpha, \frac{1}{2})
  =
  \int P(\frac{1}{2}, d\alpha)
  \,\eta(\frac{1}{2}, \alpha)
  $
  for all $\eta$, which is equivalent to
  \begin{equation*}
    P\Big(\frac{1}{2}, d\alpha\Big)
    =
    \delta\Big(\frac{1}{2}, d\alpha\Big)
    \;.
  \end{equation*}

  Similarly, in the latter case
  \begin{align*}
    \MarkovChainEnergySiteSumdist(d\sigma)
    &=
    \frac{d\sigma}{\epsilon}
    \,\Big[
    \frac{ \sigma }{\epsilon}
    \Big]^{d-1}
    \,\frac{e^{- \frac{ \sigma }{\epsilon}}
    }{ \Gamma(d) }
    \;,\qquad
    \MarkovChainEnergySiteRatiodist(d\beta)
    =
    d\beta
    \,[ \beta\,(1-\beta) ]^{\frac{d}{2}-1}
    \,\frac{ \Gamma(d ) }{ \Gamma(\frac{d}{2} )^2 }
  \end{align*}
  follows for
  $\MarkovChainEnergySiteSumdist$,
  $\MarkovChainEnergySiteRatiodist$,
  where we used the well-known properties of
  Gamma and Beta distributions.
  The reversibility condition \eqref{eqn_reversibility_s_beta}
  becomes
  \begin{align*}
    \int_0^1
    d\beta
    &
    \,[ \beta\,(1-\beta) ]^{\frac{d}{2}-1}
    \,\frac{ \Gamma(d ) }{ \Gamma(\frac{d}{2} )^2 }
    \,\rateRatio(\beta)
    \int P(\beta,d\alpha)
    \,\eta(\alpha, \beta)
    \\
    &=
    \int_0^1
    d\beta
    \,[ \beta\,(1-\beta) ]^{\frac{d}{2}-1}
    \,\frac{ \Gamma(d ) }{ \Gamma(\frac{d}{2} )^2 }
    \,\rateRatio(\beta)
    \int P(\beta, d\alpha)
    \,\eta(\beta, \alpha)
  \end{align*}
  for all $\eta$, and
  \begin{align*}
    p(d\beta)
    &=
    d\beta
    \,[ \beta\,(1-\beta) ]^{\frac{d}{2}-1}
    \,\frac{ \Gamma(d ) }{ \Gamma(\frac{d}{2} )^2 }
    \,\rateRatio(\beta)
    \,\frac{1}{Z}
  \end{align*}
  must be the expression for the unique stationary distribution of $P$.

  This proves that if the product measure is reversible, then
  $\MarkovChainEnergySitedist$ is either constant, or a Gamma distribution,
  and the transition kernel must have the claimed stationary
  distribution.

  To finish the proof it remains to consider the converse.
  Assume either of the two possible distributions for
  $\MarkovChainEnergySitedist$ and also the corresponding
  assumption on $P$. For these special distributions it is
  well-known (and easily verified) that
  the sum and the ratio are independent with the distributions
  as considered above. Hence we see that the reversibility
  condition \eqref{eqn_reversibility_s_beta} is indeed satisfied.
\end{proof}

We finish the discussion of reversible product measures
with the following remark.
Note that in the statement of
\Thmref{thm_reversibleProductMeasures}
there is an assumption on the kernel $P$ that appears
in the generator of the process $\MarkovChainEnergyT{t}$.
By \Lemref{lem_reversibleProductMeasures_systemSize}
and
\Corref{cor_reversibleProductMeasures_rates}
it suffices to consider the reversibility of the product measure
for $N=2$ and constant rates. Upon restricting this process to
any of the invariant sets $\stateSpace_{\epsilon,2}$,
the embedded discrete time
Markov chain is precisely the Markov chain on $[0,1]$ with
transition kernel $P(\beta, d\alpha)$. Therefore, the assumption
in \Thmref{thm_reversibleProductMeasures}
on the kernel $P$ is equivalent
to saying that for $N=2$ and constant rates the process
$\MarkovChainEnergyT{t}$ has a unique stationary distribution on
any of the $\stateSpace_{\epsilon,2}$.

A sufficient condition for this uniqueness is to assume that $P$
satisfies a uniform minorization condition, i.e.
there exists a constant $\gamma>0$ and a probability measure
$P^\star$ on $[0,1]$ with
$\int P^\star(d\alpha)\,\alpha = \frac{1}{2}$ and $\varPstar<\frac{1}{4}$
such that $P(\beta, .) \geq \gamma\,P^\star(.)$ for all $\beta \in [0,1]$.
Recall that this is also the type of condition $P$ assumed in
\Thmref{thm_comparisionTheorem}.

\section{Example: the rarely interacting billiard lattice}
\label{section_examples}

Here we illustrate
the use of \Thmref{thm_comparisionTheorem} and
\Thmref{thm_reversibleProductMeasures}
with the billiard lattice model studied in
\cite{10863485}, which was one of the main motivations
for our work presented in this paper.
It was argued in \cite{10863485} that in the limit of rare collisions
the dynamics of a billiard lattice becomes a Markov jump process.
The notation used in \cite{10863485} differs from ours
in that we separate the rate
of interaction $\rateSum\,\rateRatio$
from the transition probability kernel $P$, whereas in
\cite{10863485} the product $\rateSum\,\rateRatio\,P$
is denoted by $W$, and the rate function $\rateSum\,\rateRatio$
is denoted by $\nu$.
Changing equations $(61)$ and $(62)$ of
\cite{10863485}
to our notation yields
\begin{align*}
  \frac{P(\beta,d\alpha)}{d\alpha}
  &=
  \frac{3}{2}
  \,\frac{
  1 \Min \sqrt{ \frac{ \alpha \Min (1-\alpha) }{ \beta \Min (1-\beta) } }
  }{ \frac{1}{2} + \beta \Max (1-\beta) }
  \;,\quad
  \rateRatio(\beta)
  =
  \frac{\sqrt{2\pi}}{6}
  \,\frac{ \frac{1}{2} + \beta \Max (1-\beta) }{\sqrt{\beta \Max (1-\beta)}}
  \;,\quad
  \rateSum(s) = \sqrt{s}
\end{align*}
for the transition kernel $P$ and the rate functions $\rateSum$ and
$\rateRatio$, respectively. The symbol $\Max$ denotes the maximum
and $\Min$ denotes the minimum.

Since the underlying mechanical model has a three-dimensional
configuration space for each of the constituent particles it
follows that
\begin{align*}
  d
  &=
  3
  \;,\qquad
  \MarkovChainEnergySitedist(d\EnergyStateI{1})
  =
  \frac{d\EnergyStateI{1}}{\epsilon}
  \,\sqrt{ \frac{ \EnergyStateI{1} }{\epsilon} }
  \,\frac{2 \,e^{- \frac{ \EnergyStateI{1} }{\epsilon}}
  }{\sqrt{\pi}}
  \\
  \MarkovChainEnergySiteSumdist(d\sigma)
  &=
  \frac{d\sigma}{\epsilon}
  \,\Big[
  \frac{ \sigma }{\epsilon}
  \Big]^2
  \,\frac{e^{- \frac{\sigma}{\epsilon}}
  }{2}
  \;,\qquad
  \MarkovChainEnergySiteRatiodist(d\beta)
  =
  d\beta
  \,\sqrt{\beta\,(1-\beta)}
  \,\frac{8}{\pi}
  \\
  p(d\alpha)
  &=
  d\alpha\,\sqrt{\alpha\,(1-\alpha)}\,\frac{8}{\pi}
  \,\rateRatio(\alpha)\,\frac{1}{Z}
\end{align*}
should be used in
\Thmref{thm_reversibleProductMeasures}.
In fact, this measure is the (canonical)
Gibbs measure for the mechanical model, and thus must also be invariant
for the limiting jump process.

Another general property that the jump process inherits from
the underlying mechanical model is that the rate function
$\rateFunction$ is proportional to the square root of the
total energy of the two sites that interact, i.e.
$\rateSum(\sigma) = \sqrt{\sigma}$ as mentioned above. This cannot be avoided
when taking scaling limits of interacting mechanical models, because
it corresponds to the kinematic scaling relation between the energy and
the velocity (and hence the time scale). However, a rate function without
a uniform lower bound
leads to serious technical complications at various levels.
See, for example, \cite{2010arXiv1010.3972D} for
how this issue seriously complicates the rigorous derivation
of the weak interaction limit of a related deterministic model.

Furthermore, such a rate function also complicates the rigorous analysis
of the rate of convergence to equilibrium. In fact,
in order to apply the results established in this paper
we need to have $\rateSum$ bounded from below.
Recall that we showed in
\Lemref{lem_reversibleProductMeasures_systemSize}
that the above reversible product measure is also a reversible
stationary distribution for the process generated by the
infinitesimal generator corresponding to any other function $\rateSum$
(while keeping $\rateRatio$ and $\transitionP$ unchanged).
And since $\rateSum$ represents the kinematic scaling, and not the
nature of the energy exchange during an interaction,
we will change the model of
\cite{10863485}
in that we change $\rateSum$. In fact, our next Lemma is most useful exactly under the setup of the aforementioned work.

\begin{lemma}
  \label{lem_billiardNetwork}
  If $\rateSum$ is replaced by any non-negative continuous
  function, which is bounded away from zero, then the following
  hold for any $N$ and $\epsilon$.
  \begin{enumerate}
    \item
      The product measure
      $
      \MarkovChainEnergydist(d\EnergyState)
      =
      \MarkovChainEnergySitedist(d\EnergyStateI{1})
      \cdots
      \MarkovChainEnergySitedist(d\EnergyStateI{N})
      $
      with
      $
      \MarkovChainEnergySitedist(d\EnergyStateI{1})
      =
      \frac{d\EnergyStateI{1}}{\epsilon}
      \,\sqrt{ \frac{ \EnergyStateI{1} }{\epsilon} }
      \,\frac{2 \,e^{- \frac{ \EnergyStateI{1} }{\epsilon}}
      }{\sqrt{\pi}}
      $
      is the unique reversible product measure for $\MarkovChainEnergyT{t}$.
    \item
      On every $\stateSpace_{\epsilon,N}$ there exists a unique
      stationary distribution
      $
      \stationaryDistEN{\epsilon}{N}
      $. This measure is obtained by conditioning
      $\MarkovChainEnergydist(d\EnergyState)$.
    \item
      The spectrum $\spectrum{\MarkovGenerator}$
      of the generator $\MarkovGenerator$ acting
      on $\LpMu{2}{\stationaryDistEN{\epsilon}{N}}$
      satisfies
      \begin{equation*}
        \spectrum{\MarkovGenerator}
        \subset
        \Big(-\infty,
        -
        C\,\sin^2\Big[ \frac{\pi}{N+2} \Big]
        \Big]
        \cup \{0\}
      \end{equation*}
      for some constant $C$, which may depend on the choice of
      $\rateSum$.
  \end{enumerate}
\end{lemma}
\begin{proof}
  The explicit expressions for the transition kernel and the rate functions
  allow us to show
  \begin{align*}
    \rateRatio(\beta)
    \,\frac{P(\beta,d\alpha)}{d\alpha}
    &=
    \frac{\sqrt{2\pi}}{4}
    \,\frac{
    1 \Min \sqrt{ \frac{ \alpha \Min (1-\alpha) }{ \beta \Min (1-\beta) } }
    }{ \sqrt{\beta \Max (1-\beta)} }
    =
    \frac{\sqrt{2\pi}}{4}
    \,\frac{
    \sqrt{\beta \Min (1-\beta)} \Min \sqrt{\alpha \Min (1-\alpha)}
    }{ \sqrt{\beta\,(1-\beta)} }
  \end{align*}
  for all $\alpha,\beta \in [0,1]$. Hence
  $
  p(d\beta)\,P(\beta,d\alpha)
  =
  p(d\alpha)\,P(\alpha,d\beta)
  $, i.e.
  \begin{align*}
    \int p(d\beta) \int P(\beta,d\alpha) \,\psi(\alpha, \beta)
    &=
    \int p(d\beta) \int P(\beta, d\alpha) \,\psi(\beta, \alpha)
  \end{align*}
  holds for all $\psi \from [0,1]^2 \to \reals$.

  Furthermore, the estimate
  \begin{align*}
    \frac{P(\beta,d\alpha)}{ \MarkovChainEnergySiteRatiodist(d\alpha) }
    &=
    \frac{3\pi}{16}
    \,\frac{
    1 \Min \sqrt{ \frac{ \alpha \Min (1-\alpha) }{ \beta \Min (1-\beta) } }
    }{ \frac{1}{2} + \beta \Max (1-\beta) }
    \,\frac{1}{\sqrt{\alpha\,(1-\alpha)}}
    \\
    &=
    \frac{3\pi}{16}
    \,\Big[
    \frac{
    \frac{ 1 }{\sqrt{\alpha\,(1-\alpha)}}
    }{ \frac{1}{2} + \beta \Max (1-\beta) }
    \Min
    \frac{
    \frac{1}{\sqrt{\alpha\Max (1-\alpha)}}
    }{ [\frac{1}{2} + \beta \Max (1-\beta)]\, \sqrt{\beta \Min (1-\beta)} }
    \Big]
    \\
    &\geq
    \frac{3\pi}{16}
    \,\Big[ \frac{4}{3} \Min \sqrt{2} \Big]
    =
    \frac{\pi}{4}
  \end{align*}
  provides the minorization condition
  $
  P(\beta,d\alpha)\geq
  \frac{\pi}{4}\,\MarkovChainEnergySiteRatiodist(d\alpha)
  $.
  In particular, this implies that the Markov chain on $[0,1]$
  with transition kernel $P$ has a unique invariant measure.

  Therefore,
  it follows then from \Thmref{thm_reversibleProductMeasures}
  that $\MarkovChainEnergydist(d\EnergyState)$
  is a reversible product measure, and must be unique.

  Observe that by
  \Thmref{thm_reversibleProductMeasures}
  the infinitesimal generator corresponding to
  $ P^\star(d\alpha) = \MarkovChainEnergySiteRatiodist(d\alpha) $
  and a constant rate function
  also has
  $\stationaryDistEN{\epsilon}{N}$ as a stationary reversible
  distribution.
  Combining this with the above minorization condition for
  $\transitionP$
  and
  $
  \frac{\sqrt{\pi}}{3}
  \leq
  \rateRatio(\beta)
  \leq
  \frac{\sqrt{2\pi}}{4}
  $
  we see that
  under the assumption that $\rateSum$ is bounded from below
  all assumptions of
  \Thmref{thm_comparisionTheorem} are satisfied.
\end{proof}

The significance of the result of
\Lemref{lem_billiardNetwork}
is that it provides an interesting model that fits the
conditions of
\Thmref{thm_comparisionTheorem}.
We would like to point out that
previous to
\cite{10863485}
the analogous two-dimensional billiard network
was studied in
\cite{10376859}.
However, in this case the uniform mixing condition
\eqref{condition_mixing_P}
of
\Thmref{thm_comparisionTheorem}
fails to hold, which is why we restricted our attention in the above to the
three-dimensional setting.

\section{Conclusion}

\label{sect_conclusion}

The authors of \cite{10863485} suggested a two step strategy for deriving
the heat equation from a mechanical model.
Motivated by that we have introduced in this work a class of stochastic
models with the aim to implement the second step of their strategy:
the derivation of the heat equation from a mesoscopic stochastic model.

At present it is widely understood that a necessary ingredient
to rigorously establish the hydrodynamical limit
is a sharp bound on the dependence of the spectral gap of the generator
on the system size. Such a bound is one of the main results of the present
paper.

Besides the importance of this bound for the hydrodynamic limit, an additional
value of our result is that for systems with continuous state space
such bounds are hard and scarce, e.g. \cite{MR1825150,MR2020418}.
As in those works, our method requires to assume that the rates
are bounded away from zero.

In more detail: according to our main result the spectral gap of the
infinitesimal generator of the process
scales as $\Order(N^{-2})$ in terms of the systems size $N$.
This is precisely the kind of scaling which allows for a
diffusive scaling limit,
and hence the study of the hydrodynamic limit.
However, we do not study the hydrodynamic limit in this paper,
because it requires different ideas and techniques, and results on
the spectral gap are of interest in their own right.

To keep the model as close to the mechanical ones as possible
(cf. \Sectionref{section_examples} and \cite{10863485})
it is desirable to remove the assumption on existence of a
uniform lower bound of the rate function. Numerical simulations
suggest that the $\Order(N^{-2})$ scaling of the spectral gap
remains true also for
rate functions that can approach zero. In particular for the square root
of the total energy of the interacting pair, which is the rate function
that appears in mechanical models due to the kinematic scaling of
the velocity with the energy. However, we do not have a rigorous proof
of such a statement available at present.

\end{document}